\documentclass{statsoc}
\RequirePackage{natbib}

\usepackage{xspace}
\usepackage{amsmath,  amssymb}
\usepackage{fancyhdr, fancybox, graphicx,psfrag,dcolumn,bm,accents, textcomp,natbib}
\usepackage[colorlinks=true, urlcolor=blue, citecolor=blue, linkcolor=blue]{hyperref}
\usepackage{color}
\usepackage{stmaryrd}
\usepackage[mathlines]{lineno}
\usepackage[usenames,dvipsnames]{xcolor}

\graphicspath{{Pics/}}

\newcommand{\lvec}[1]{\overset{{}_{\mbox{\hspace{.5mm}}\shortleftarrow}}{#1}}
\newcommand{\rvec}[1]{\overset{{}_{\shortrightarrow}}{#1}}

\newcommand{\matY}{\mathbf{Y}}
\newcommand{\vecmu}{\rvec{\mu}}
\newcommand{\vecalpha}{\rvec{\alpha}}
\newcommand{\vecP}{\overset{{}_{\mbox{\hspace{.5mm}}\shortrightarrow}}{P}}
\newcommand{\vecR}{\lvec{R}}
\newcommand{\vecS}{\lvec{S}}
\newcommand{\veca}{\rvec{a}}

\newcommand{\vecm}{\rvec{m}}

\newcommand{\matEta}{\boldsymbol{\eta}}

\newcommand{\vecEtaj}{\rvec{\eta}_{,j}}
\newcommand{\etagj}{\eta_{g,j}}
\newcommand{\etagij}{\eta_{g(i),j}}

\newcommand{\matEtaNotj}{\boldsymbol{\eta}_{,/j}}
\newcommand{\tissuegene}{tissue-gene\xspace}
\newcommand{\tissuegenes}{tissue-genes\xspace}

\newcommand{\dayfive}{day 5\xspace}
\newcommand{\parentoforigin}{parent-of-origin\xspace}
\newcommand{\Parentoforigin}{Parent-of-origin\xspace}
\newcommand{\X}{\emph{X}\xspace}

\newcommand{\Xchromosome}{\emph{X}-chromosome\xspace}

\newcommand{\Xinactivation}{\emph{X}-inactivation\xspace}

\newcommand{\Xlinked}{\emph{X}-linked\xspace}
\newcommand{\Xlocated}{\emph{X}-located\xspace}

\newcommand{\RNAseq}{RNA-seq\xspace}
\newcommand{\weightbiasedcoin}{weight-biased coin\xspace}
\newcommand{\WeightBiasedCoin}{Weight-Biased Coin\xspace}

\newcommand{\preexisting}{preexisting\xspace}
\newcommand{\deactivate}{inactivate\xspace}

\newcommand{\FOne}{$\text{F}_1$\xspace}

\newcommand{\Ftwo}{$\text{F}_2$\xspace}
\newcommand{\Xce}{\emph{Xce}\xspace}
\newcommand{\Xist}{\emph{Xist}\xspace}
\newcommand{\Xcea}{$\mbox{\emph{Xce}}^{a}$\xspace}
\newcommand{\Xceb}{$\mbox{\emph{Xce}}^{b}$\xspace}
\newcommand{\Xcec}{$\mbox{\emph{Xce}}^{c}$\xspace}

\newcommand{\StrainName}[1]{\texttt{#1}\xspace}

\newcommand{\OneTwoNine}{\StrainName{129}}
\newcommand{\AJ}{\StrainName{AJ}}
\newcommand{\ALS}{\StrainName{ALS}}
\newcommand{\BSix}{\StrainName{B6}}
\newcommand{\CAST}{\StrainName{CAST}}
\newcommand{\LEWES}{\StrainName{LEWES}}

\newcommand{\PWK}{\StrainName{PWK}}
\newcommand{\SJL}{\StrainName{SJL}}
\newcommand{\WLA}{\StrainName{WLA}}
\newcommand{\WSB}{\StrainName{WSB}}

\newcommand{\TIRANO}{\StrainName{TIRANO}}
\newcommand{\ZALENDE}{\StrainName{ZALENDE}}
\newcommand{\PERA}{\StrainName{PERA}}
\newcommand{\DDK}{\StrainName{DDK}}
\newcommand{\PANCEVO}{\StrainName{PANCEVO}}

 \usepackage[T1]{fontenc}

\title[Beta model for X-inactivation]{Bayesian Manifold-Constrained-Prior Model for an Experiment to Locate \emph{Xce}}

\author[Alan Lenarcic {\it et al.}]{Alan B. Lenarcic}\email{lenarcic@post.harvard.edu}
\address{Department of Genetics, University of North Carolina at Chapel Hill, USA.}

\author{John D. Calaway}\email{Calaway.jdc@gmail.com}
\address{Department of Genetics, University of North Carolina at Chapel Hill, USA.}
 
\author{Fernando Pardo-Manuel de Villena}\email{fernando@med.unc.edu}
\address{Department of Genetics, University of North Carolina at Chapel Hill, USA.}
	
\author{William Valdar}\email{william.valdar@unc.edu}
\address{Department of Genetics, University of North Carolina at Chapel Hill, USA.}

\begin{document}
\bibliographystyle{chicago}
\begin{abstract}
  	We propose an analysis for a novel experiment intended to locate the genetic locus \Xce (X-chromosome controlling element), which biases the stochastic process of \Xinactivation in the mouse. \Xinactivation bias is a phenomenon where cells in the embryo randomly choose one parental chromosome to inactivate, but show an average bias towards one parental strain. Measurement of allele-specific gene-expression through pyrosequencing was conducted on mouse crosses of an uncharacterized parent with known carriers.  Our Bayesian analysis is suitable for this adaptive experimental design, accounting for the biases and differences in precision among genes.  Model identifiability is facilitated by priors constrained to a manifold. We show that reparameterized slice-sampling can suitably tackle a general class of constrained priors.  We demonstrate a physical model, based upon a ``weighted-coin'' hypothesis, that predicts \Xinactivation ratios in untested crosses.  This model suggests that \Xce alleles differ due to a process known as copy number variation, where stronger \Xce alleles are shorter sequences.
          \end{abstract}

\section{Introduction}\label{Section1:Introduction}
\begin{figure}[htbp]
	\begin{center}
		\includegraphics[width=1.0\textwidth]{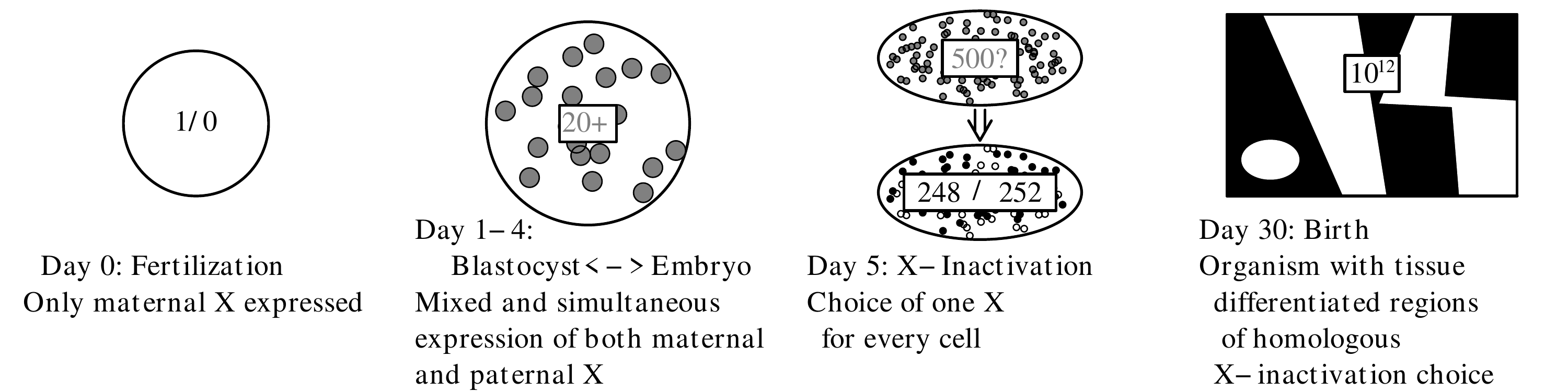}
		\end{center}
	\caption{Schematic of \Xinactivation in embryonic development~\citep{LYON1961}.  Near \dayfive~\citep{Mak30012004,Takagi1982},  progenitor cells \deactivate either their maternal or paternal \Xchromosome, turning it into a Barr Body\citep{Barr1949, Lee2011}, later passing this choice to future daughter cells.  
	The full organism at birth is a mosaic of regions of predominant maternal or paternal expression, under usual circumstances expected at an overall 50-50 balance.  This balance is disturbed if there is a survival disadvantage for cells expressing either \Xchromosome, or if there was an \Xce bias influencing the initial selection.  In this paper we measure \Xlinked gene expression on multiple mouse pup tissues, with one goal of inferring the \Xinactivation proportions initially set on \dayfive.}
	\label{fig:ARoughFlowchart}
\end{figure}

   \sloppy \Xinactivation bias is a phenomenon observed in the female mouse, first proposed in~\cite{Cattanach1967} to be caused by a gene \Xce (X-controlling element), which has unknown location and mechanism.  In mammals, cells of the early female (XX) embryo inactivate one of their parental chromosomes through a random choice, and pass this decision onto their daughter cells~\citep{Lyon1962,Gendrel2011}, as shown in Figure~\ref{fig:ARoughFlowchart}.  Thus, in mammals, tissues have mosaic regions of parental chromosome preference; this can be seen, for example, in the fur of a calico cat, whose two fur colors reflect its two parents.  \cite{Cattanach1965, Cattanach1967} first discovered an \Xinactivation bias in mice by shaving mice with parents of white and black hair follicle color, counting the ratio of colors in the hybrid animal, and observing an average bias toward one color.  In our related report, \cite{JohnCalaway2013}, we proposed to investigate for the location of \Xce, refining an interval proposed in~\cite{Chadwick2006}, by ascertaining the allele carried by 10 unclassified strains.  For experimental reasons relating to cost and accuracy, that study measured allele-specific expression in infant hybrids using the technology of pyrosequencing.

   Here we provide a model for these data, specific to the needs of the experiment, including the adaptive nature of the experimental design.  We propose a Bayesian MCMC-based hierarchical model, based upon the beta-distribution. The model relates allele-specific proportions to a summary proportion reflecting the initial proportion of progenitor cells that conducted the initial choice around 5 days in the embryo development.  Our use of the beta distribution suggests a P\`{o}lya-urn model, which can estimate the number of cells in each organ at \Xinactivation. We use posterior tail values to argue for the identity of the allele carried by an unknown strain, based on its average bias against known allele-carriers.
	
	To make our model identifiable, we propose priors for sets of our parameters that are constrained to a manifold condition.  Through reparameterizing the slice-sample distribution, we can generate acceptable draws from the correct, constrained posterior.  Constrained priors become especially necessary when we attempt to design a physical model for the \Xinactivation bias observed in the crosses.  To predict \Xinactivation bias in potential crosses between alleles, we fit a ``weight-biased coin'' physical model.   This physical model supports the hypothesis that \Xce is a genetic locus based upon copy-number variation~\citep{Stankiewicz2010, Sebat2004}.  Our simulations and analysis here strengthen the case for the findings of the ~\cite{JohnCalaway2013} \Xce experiment and  
demonstrate value for Bayesian hierarchical modeling in future allele-specific-expression experiments.

\section{Allele-specific Expression Experiment for Locating \emph{Xce}}\label{Section2:ExperimentalMethodAndBackground}
  Prior to this experiment, four alleles ($a,b,c,e$) of \Xce were known in mice, found, for example, in standard strains of mice \AJ: $a$,  C657B6/J (\BSix): $b$,  CAST/EiJ(\CAST): $c$, and in SPRET/EiJ, of the separate mouse species \emph{Spretus}:$e$~\citep{Cattanach1967, BM.1994,C.1991}.  We have related these alleles with respect to the order of their ``strength''.  An \AJ$\times$\BSix hybrid which is $a\times b$ will favor expressing the \BSix \Xchromosome on average in approximately 65\% of their cells;  an $a\times c$ cross produces approximately a 30-70 ratio; for $a\times e$ it is 20-80.  	  Even when the population average for a given hybrid is 50-50, there is considerable variation in individual ``skew'', i.e. the proportion an individual might over-express one parental \Xchromosome. Skewed ratios of 75-25, 27-72, etc. are observed in individuals, and the allele specific expression individual genes might present additional imprecision or biases.

  \cite{Chadwick2006} proposed a location of \Xce to somewhere within a $\sim$1.9 Mb (million base-pairs) region on the \Xchromosome.  
	  This was based upon identification of the allele carried by strains of mice, originally conducted with techniques like shaving and counting hair follicles of hybrids with differently colored parents.  
		But counting hair follicles imprecisely measures \Xinactivation only in mice old enough to be shaven and only in skin tissues.  Given the costs involved in studying embryos and laboratory expertise,~\cite{JohnCalaway2013} sought an efficient and reliable measure of aggregate whole-body expression taken from newborn tissue.
	  Due to variation of \Xinactivation skew among clones of
		identical genome, multiple individuals must be 
		measured to establish that a given ``first-generation hybrid'' (\FOne) cross shows a bias, favoring measurement techniques of low cost. 
		
		Pyrosequencing, a ``sequencing by synthesis method'' introduced in~\cite{Ronaghi1998}, was viewed as a cost-effective technology to measure gene expression for strains whose genome sequences were known but whose \Xce alleles were not.  \cite{JohnCalaway2013} chose commonly expressed \Xlocated genes: \emph{Ddx26b}, \emph{Fgd1}, \emph{Rragb}, \emph{Mid2},  \emph{Zfp185}, \emph{Xist},  \emph{Hprt1}, and \emph{Acsl4}, which featured known SNPs that differentiate two strains. 
   Pyrosequencing reads target-sequences of length 20-30 base pairs including either allele of a SNP:  Exponential range PCR amplifies up to 1 micro-liter ($\mu l$) of the target, or approximately $3.7 \times 10^{10}$ copies, which, after purification, are counted through synthesis.  Pyrosequencers report measurements in proprietary fluorescence units, where, conservatively, the final ratio of allele-specific gene expression is based upon more than $1$ million targeted reads.

\begin{table}
\caption{Information on common inbred mouse strains used in this experiment.}	
	\centering
 \begin{tabular}{l|c|llll}
 Strain & Abbrev. & details & Sub-species & derived &  \Xce allele \\ \hline 
 A/J  &  \AJ & Albino (hearing loss) & \emph{domesticus} & lab & $a$ \\
 129S1/SvlmJ & \OneTwoNine & Brown (testicular cancer) & \emph{dom.} & lab & $a$ \\
 C57BL/6J & \BSix & ``Black-six'' & \emph{dom.} & lab & $b$ \\
 ALS/LtJ  & \ALS &  Lou Gehrig's disease& \emph{dom.} & lab & $b$ \\
 CAST/EiJ & \CAST & & \emph{castaneus} & wild & $c$ \\
 WSB/EiJ  &  \WSB & ``Watkin Star'', aka ``White Spot''& \emph{domesticus} & wild & ?? \\
 WLA/Pas & \WLA & Malaria Resistant & \emph{dom.} & wild & ?? \\
 LEWES/EiJ & \LEWES  &  Caught in Lewes, DE & \emph{dom.} & wild & ??\\
 PWK/PhJ  & \PWK &    & \emph{musculus} & wild & ?? \\
 SJL/J & \SJL  & Albino (sight loss) & \emph{dom.} & lab & ??\\
 TIRANO/EiJ & \TIRANO&  Poschiavinus Valley, Tirano, Italy& \emph{dom.} & wild & ?? \\
 ZALENDE/Ei & \ZALENDE & Zalende, Switzerland & \emph{dom.} & wild & ?? \\
 DDK/Pas &\DDK & Fertility loss, Japanese & \emph{dom.} & lab & $a$ \\
 PERA/EiJ & \PERA & Rimac Valley, Peru & \emph{dom.} & wild & ?? \\
 PANCEVO/EiJ& \PANCEVO & Pancevo, Serbia & \emph{spicilegus} & wild & ?? \\
   \end{tabular}
   
	  \label{Table:NamesOfMice}
\end{table}

		Table~\ref{Table:NamesOfMice} gives a short reference background to the mouse types used in this experiment.
 There are three mouse sub-species: \emph{mus musculus domesticus} from Europe, \emph{mus musculus castaneus} from Africa, India and the Arabian subcontinent, and \emph{mus musculus musculus} from Asia, that rarely interbreed in the wild (\PANCEVO is a different species: \emph{mus spicilegus}, ``Steppe Mouse'').  Laboratory-derived strains were initially generated from domesticated toy mice.  Wild-derived strains come from wild-captured samples, inbred to be representatives of a wild subpopulation.  We wish to classify  \Xce allele membership for strains marked with unknown allele ``??'', and thus narrow down a candidate region for \Xce.

\subsection{Implied Sequential Design of the Experiment}

The~\cite{JohnCalaway2013} first established which \Xce allele was carried by a few unclassified strains.   Based upon the observations,  a provisional, semi-statistical classification was assumed, and this motivated the choice to classify additional strains.  First, strains  SJL/J (\SJL) and ALS/LtJ (\ALS) were analyzed because their sequences split the known region between \Xcea and \Xceb.  Then strains LEWES/EiJ (\LEWES) and WLA/Pas (\WLA) followed.  Additional crosses with WSB/EiJ (\WSB), PWK/PhJ (\PWK) were generated and measured, including some \RNAseq measurements.  Available frozen \Ftwo samples of crosses including  
	  DDK/Pas (\DDK),  ZALENDE/Ei  (\ZALENDE), PERA/EiJ (\PERA)~\citep{Nishioka1995} were acquired.  Some successful breedings with PANCEVO/EiJ (\PANCEVO)~\citep{Kim2005} were conducted to see if this second species of mouse had a different \emph{Xce} profile.
 Only a subset of possible crosses were performed, and not all reciprocal crosses were generated.  Count of replicates per unique crosses varied from 4 to 50.

	For convenience,  summarizing details for the inbred parental strains are  given in Table~\ref{Table:NamesOfMice}.
A statistical analysis must respect that the experiment is adaptive and sequential in nature, that many crosses which might have aided in the statistical estimate will be missing, and that we should provide a statistical criterion not only to assert that a given cross \SJL$\times\BSix$ has an average ``$50.1\% \pm 2.1\%$'', but that we should also be able to express how certain we are that $\SJL$ is an $a$ allele carrier.  By choosing closely related potential \Xcea and \Xceb  allele carriers, predominantly from \emph{mus domesticus}, ~\cite{JohnCalaway2013} proceeded with an investigation that, aided by this statistical analysis, refined the \Xce location from the 1.9 Mega-base (Mb) \cite{Chadwick2006} candidate interval to a 176 kilo-base (kb) region approximately 500 kb from the gene \Xist whose expression causes \Xinactivation.   \cite{JohnCalaway2013} established that allele \Xcea is rare in the wild and that most individuals within a sub-species share the same \Xce allele.  They found that \Xce-allele strengths can potentially increase or decrease during speciation, which suggests \Xce is mutated through copy number variation.

\subsection{Format of Pyrosequencing Data}

	       A partial Table~\ref{Table:ExamplePartTable} shows the challenges of the~\cite{JohnCalaway2013} data.  Every mouse is a member of a single cross, which is coded to correspond to a unique hybrid combination of mother and father inbred strains.   We encode every cross with a short code, such as ``1Wl'' for a \OneTwoNine$\times$\WLA cross, where \OneTwoNine is the maternal strain.  Note that a \WLA$\times$\OneTwoNine hybrid, with \WLA as maternal strain, would be coded ``Wl1'' and coded as a different reciprocal cross.  The other columns are measurements of maternal proportion expressed for given \tissuegene combinations.
		Allele-specific proportion is measured in multiple \tissuegenes, with considerable missingness: ``NA'' for genes that were either dropped at random for an individual mouse during experiment, or because they are not experimentally viable for the cross when no SNP exists on that gene  to differentiate maternal and paternal copies.  Since SNPs are single base-pair substitutions, between any three parental strains there is no single SNP that can be used to measure allele-specific expression for all three pairings of two strains.

   \begin{table}
	\caption{Example section of pyrosequencing maternal expression ratios dataset.}
	\centering   \begin{tabular}{c|ccc|ccccc}
pup & cross & dam & sire  & $\begin{array}{c} \mbox{Brain} \\ \mbox{Ddx26b} \end{array}$	&	
 $\begin{array}{c} \mbox{Brain} \\ \mbox{Rragb} \end{array}$	
 &	$\begin{array}{c} \mbox{Kidney} \\ \mbox{Ddx26b} \end{array}$ & 
 $\begin{array}{c} \mbox{Kidney} \\ \mbox{Rragb} \end{array}$  &\ldots \\
\hline
3-7	&1Wl	& 129S1-3	& WLA-3	&	0.234	&NA	&0.183 &  NA & \ldots\\
5-15&	1Wl	&129S1-5	&WLA-5	&	0.316 & NA & 
	NA	& NA & \\ 
	\ldots & \ldots & & & & & \\
2-1 &	1Wl	& 129S1-2	&WLA-7	&	0.606	&	NA&	0.483	& NA &\\
7-4	& 1Wl	& 129S1-7	&WLA-7	&	0.648		&NA	&NA	 & NA & \\
16-3 &	AlAj	& ALS-16 &	AJ-873	&	0.456	 &	.501	&0.447 & .51&	\\
16-5 &	AlAj	& ALS-16 &	AJ-873	&	0.574	&	.432	&0.532 & .59 &	\ldots \\
\ldots
\end{tabular}  \label{Table:ExamplePartTable} \end{table}

  If we naively take sample means of all observed gene expression in individuals stemming from one type of cross, we can summarize those averages in a table like Table~\ref{Tab:Sum1}, where we see that of candidates with historically established alleles $a$, $b$, $c$, there are similar behaviors encountered in crosses with previously unknown \PWK, \SJL, \WSB strains.  Reciprocal crosses, such as the $b\times a$ cross, have often not been measured.  We inevitably hope for a statistical method that can establish the allele carried by unestablished strains. 
  
  Allele-specific expression could have been measured through Illumina \RNAseq~\citep{Morin2008} or microarrays~\citep{Chang1983, Hall2007}. \RNAseq  whole-transcriptome sequencing  enables simultaneous measurement of allele-specific expression of all 473 \Xlocated genes.   But typical \RNAseq will only have on order 500 read counts of any SNP region, offering less precision than pyrosequencing.  A Bayesian method for estimating precision and bias with \RNAseq count data can be found in~\cite{Graze2012}.  Both \RNAseq and pyrosequencing are unbiased at preserving true, initial
allelic stoichiometry.   But the non-linear relationship between expression and luminescence in microarrays can lead to biased estimation of allele-specific proportion.       \begin{table}
		\caption{Unweighted sample mean expression, maternal allele (rows), paternal allele (columns).}
		\centering
     
		$
    \begin{array}{c|cccccc} & a & b & c & \mbox{??}_{L}  & \mbox{??}_{P}  & \mbox{??}_{S}  \\ \hline 
       a &  &  &  &  & .32 &   \\ 
       b & .57 & .51 & .33 &  & .54 &   \\ 
       c &  &  &  &  & .75 &   \\ 
       \mbox{??}_{L}  & .59 &  &  &  &  &   \\ 
       \mbox{??}_{P}  & .7 & .58 & .36 &  &  &   \\ 
       \mbox{??}_{S}  & .56 & .48 & .34 & .5 &  &   \\ 
     \end{array} 
       \mbox{, where: } 
    \begin{array}{c|l} a & \mbox{\AJ}, \mbox{\OneTwoNine} \\ 
        b & \mbox{\BSix}, \mbox{\ALS} \\ 
        c & \mbox{\CAST} \\ 
        \mbox{??}_{L} & \mbox{\LEWES} \\ 
        \mbox{??}_{P} & \mbox{\PWK} \\ 
        \mbox{??}_{S} & \mbox{\SJL} \\ 
        \end{array} 
        $
				
				\label{Tab:Sum1} \end{table}
				
  Late in the~\cite{JohnCalaway2013} experiment, brain-tissue \RNAseq measurements were made available for 34 individuals from \FOne crosses between wild-derived sub-species representatives \WSB, \PWK, and \CAST, about $6$ individuals of each cross, including reciprocal crosses.  	It was hoped that our statistical procedure designed for pyrosequencing output could accommodate \RNAseq data.  As we show, \RNAseq measurements serve the same role in our model as pyrosequencing, although
	\RNAseq measurements often have less precision when baseline expression of a gene is low.   
  
\section{Statistical Model for Gene-expression Ratios}\label{Section3:StatisticalModel}
 \sloppy Denote the proportion of maternal gene-expression measurements $Y_{ij}\in[0.001,.999]$, where $i \in \{1, \ldots, n\}$ is the index of the mouse specimen used in the experiment, and $j \in \left\{1, \ldots, J\right\}$ is an index of \tissuegene combination (examples: ``kidney-\emph{Fgd1}'', ``brain-\emph{Ddx26b}'').  We are given $Y_{ij}$ as a fractional proportion of maternal expression over total expression.
		Certain genes, dependent on tissues,  could be biased or imprecise measures of true overall \Xinactivation.  For instance, the gene \emph{Fgd1}, which differentiates  \AJ and \CAST, might be biased in favor of \AJ expression within brain tissue but not in the liver.  Measuring multiple tissues in multiple genes, we hope to estimate a quantity, $P_{i}$, representing a best whole-body proportion of active maternal \Xchromosome. 
  
   Even if mouse $i$ has a whole-body proportion of $P_{i}$, replicate mice from the same cross $i', i'', \ldots$ will have different proportions $P_{i'}, P_{i''}, \ldots$.  The cross-specific mean $\mu_{g}$ denote the population mean for mice in cross/group $g \in \{1, \ldots, G\}$.   We seek then a hierarchical method that estimates $P_{i}$ from observed $Y_{ij}$, and relates $P_{i}$ for individuals $i$ in group $g$ to their mean $\mu_{g}$.  Only a subset of genes $j$ can be observed for certain crosses $g$.   Some data is missing-at-random, but much $Y_{ij}$ is systematically unmeasurable, due to shared SNPs for that gene.  
		Our inference model follows.
  
  \subsection{A Hierarchical Beta Model} 
    Our scientific objectives are served by a Bayesian hierarchical model.  Observed data are proportions $Y_{ij}$,    so a beta-distribution regression model~\citep{Ferrari2004} can model unimodal proportions. 
	For our problem, the mean and mode for the beta densities should be constrained within $(0,1)$, excluding $0$ and $1$.  Thus, we choose parameterized beta distributions of the form $\mbox{Beta}(\mathbb{A}+1, \mathbb{B}+1)$ where $\mathbb{A},\mathbb{B} \geq 0$.  We first give the parameterized mathematical model, and then describe the purpose of the parameters in the model.  Let:
    \begin{equation} \begin{split} 
     \mbox{Observed: } Y_{ij}  \sim \mbox{ Beta }(\mbox{ } P_{i} R_{j} & S_{j} e^{\etagj} +1, \mbox{ }
      (1-P_{i}) (1-R_{j})  S_{j} e^{\etagj} + 1 )\mbox{ , } \\
     \mbox{Where Latent: } \underbrace{P_{i}}_{\mbox{\tiny{Whole Body}}}\mbox{ } \sim \mbox{ }\mbox{ Beta }(\mbox{ }&\mu_{g} \alpha_{g} +1, (1-\mu_{g})\alpha_{g} +1) \mbox{ .}
      \end{split} \label{BetaHierarchicalModel} \end{equation}   
      In the above equation, we have latent, unobserved ratios $P_{i}$ for $i\in \{ 1, \ldots, n\}$ for each individual and parameter vectors $\vecR, \vecS \in \mathbb{R}^{J}$, $\vecalpha,\vecmu \in \mathbb{R}^{G}$ and a perturbation matrix $\matEta \in \mathbb{R}^{G \times J}$.  In our notation, $\vecR$ and $\vecS$ are row-vectors each of length $J$, the number of columns of Table~\ref{Table:ExamplePartTable}, while $\vecP, \vecmu, \vecalpha$ are column-vectors for the $n$ individuals or $G$ crosses.  The matrix $\matEta$ is of dimension $G \times J$.   
      
      $R_{j} \in (0,1)$ represents  \textbf{bias} for \tissuegene combination $j \in \{1, \ldots, J\}$.  The value $R_{j} = .5$ suggests that this \tissuegene is relatively unbiased as a predictor of overall \Xinactivation.  If a \tissuegene has $R_{j} = .6$, then we expect this to push $Y_{ij}$, on average for all crosses, in the direction of maternal expression. 
      
       In contrast, $S_{j} \in (0, \infty)$, represent \textbf{precision} of measurements for a given \tissuegene combination $j$.
			Larger values of $S_{j}$ imply that \tissuegene $j$ is a more precise measurement of $P_{i}$.  To add further difference in precision for certain crosses,  parameters $\etagj \in \{-\infty, \infty\}$  model deviation in precision for crosses $g$ of\tissuegene measurements $j$ as compared to other crosses.  The precision for cross $g$ on \tissuegene $j$ will be $S_{j} \times e^{\etagj}$.  Because the Beta-distribution implies a P\`{o}lya-urn model for cell proliferation, we can use measurement of $S_{j}$ to infer in some measure how many progenitor cells exist in each organ during the \Xinactivation event.  Let priors for $S_{j}$ be $\mbox{Gamma}(\chi_{S}, \xi_{S}) \propto e^{-x/\xi_{S} } x^{\chi_{S}-1}$ with prior for hyper-parameters $(\xi_{S},\chi_{S}) \sim \mbox{Gamma}(.1,.1)$.
       
       The $+1$ to both terms of $\mbox{Beta}(\mathbb{A}+1, \mathbb{B}+1)$ ensure stability of the posterior, keeping the beta density $\propto x^{\mathbb{A} + 1-1}(1-x)^{\mathbb{B} + 1- 1}$ from having infinite boundary density if $P_{i},\mu_{g} \approx 1$ or if $P_{i}, \mu_{g} \approx 0$.  Since we restrict our population of genes and individuals to those expressing between $1\%$ and $99\%$ of the maternal \X (individuals expressing more or less are presumed either males or XO females carrying only a single \Xchromosome due to a XY separation error in the gamete), the $(+1,+1)$ information has negligible effect on mid-range values and reflects our sampling choices.
      
      Hyperparameters $\mu_{g}, \alpha_{g}$ represent the average and variance for $P_{i}$ within a cross group $g$.  $\mu_{g} \in (0,1)$ reflects the average $P_{i}$ proportion of cross $g$, and is the most important parameter for establishing \Xce membership.  Nuisance parameter $\alpha_{g} \in (0, \infty)$ represents the variability of $P_{i}$ about $\mu_{g}$.    For large $\alpha_{g}$ values,  $P_{i}$ variation about $\mu_{g}$ will be less.  We assign $\vecmu$ a hierarchical prior:
    \begin{equation} \mu_{1}, \mu_{2}, \ldots, \mu_{G} \sim \mbox{ i.i.d. } \mbox{Beta}( \mu_{\mbox{\tiny{All}}} \alpha_{\mbox{\tiny{All}}} + 1,  (1-\mu_{\mbox{\tiny{All}}}) \alpha_{\mbox{\tiny{All}}} + 1 )\mbox{.} \label{MuPrior} \end{equation}
 For Equation~\ref{MuPrior}, the global mean and precision parameters, $\mu_{\mbox{\tiny{All}}}, \alpha_{\mbox{\tiny{All}}}$ are assigned weak hyper-priors, $\mbox{Unif}(0,1)$ and $\mbox{Gamma}(.1,.1)$ respectively.  For vector $\vecalpha$, which represents the spread of observed $P_{i}$ around their group mean $\mu_{g}$, we assign an i.i.d. $\mbox{Gamma}(1,1)$ exponential prior.

    \subsection{Constrained Priors}
    Due to identifiability, canonical i.i.d. priors do not suit our purposes for parameters $\vecR$ and $\matEta$.
 If all $J$ bias parameters $R_{j}$  deviated from $.5$ together, such as if $\bar{R} $$=$ $\frac{1}{J}\sum_{j} R_{j} = .6$, the model would shift $\hat{P}_{i}$ to the $\mbox{sign}(.5-\bar{R})$ direction of all $Y_{ij}$.  But a global movement of $\hat{P}_{i}$ would shift all $\hat{\mu}_{g}$.  Relative differences $\hat{\mu}_{g_1} - \hat{\mu}_{g_2}$ might be preserved, but we would not obtain a precise estimate of $\hat{\mu}_{g}$, if confounded with $\vecR$.

     We explicitly \underline{assume} (because we have no way of discovering otherwise) that \tissuegene combinations are ``on-average'' unbiased: $\sum_{j} \mbox{logit}(R_{j}) = 0$, where $\mbox{logit}(R_{j}) = \log_{e} \frac{ R_{j}}{1-R_{j}}$.  We will report $\hat{\mu}_{g}$ given an assumption that $\mbox{logit}(R_{j})$ is on average unbiased.  Since most decisions will be based upon the significance of differences $\hat{\mu}_{g'} - \hat{\mu}_{g}$ we feel that this assumption is reasonable.  		We therefore model bias with a restricted prior of the form:
          \begin{equation} R_{j} \sim \mbox{Marginally } \mbox{Beta}(u_R, u_R)\mbox{ conditional on }   \sum_{j} \log \frac{R_{j}}{1-R_{j}} = 0\mbox{ .}\label{Rprior} \end{equation}
	which we show in Section~\ref{SliceSamplingSection} is suitable for MCMC posterior exploration.
      
     Model parameters $\etagj$ represent deviation about parameter $S_{j}$.      Since  $\etagj \in \{-\infty, \infty\}$, Gaussian priors $N(0, \tau^2_{\eta})$ might seem acceptable.  Certainly, if all $\etagj$ deviated from $0$ together, then the combination product $e^{\etagj}S_{j}$ could support many values of $S_{j}$ to retain the same overall precision.  Inferentially, this weakens our estimation ability of $S_{j}$.  Instead, we define $\sum_{g} \etagj = 0$ for every $j$.  Thus $S_{j}$ represents an average precision over all groups $g$.     	The prior for $\etagj$ will be $\sum_{g} \etagj = 0$,  but that $\etagj \sim N(0, \tau^2_{\eta})$ marginally.
 
 Constrained priors limit the parameter space to a lower-dimensional manifold.  In Gaussian Gibbs-regression settings, such restrictions are commonly imposed~\citep{Gelfand1992}. \cite{Vines1996} willingly accept a 10x slowdown in iteration speed to estimate regressions with constrained parameters for a mixed effects model, and observed reduced autocorrelation in the samples.  In non-Gaussian settings, ~\cite{Zhu2011,Shi2009,Zho2010} show advances in manifold exploration and suggest new ways to explore these parameter spaces.  We demonstrate a slice-sampler suitable for our linear constraints in Section~\ref{SliceSamplingSection}.

Due to orthogonality in design between $\lvec{R}$ and $\mu_{g}$, we can approximate that for $J$ fixed and $n_{g} \rightarrow \infty$:
  \begin{equation} \sqrt{n_{g}} ( \hat{\mu}_{g} - \mu_{g}) \Longrightarrow  N\left(0,  \mu_{g}(1-\mu_{g})(\sigma^2_{u} + \sigma^2_{e}/J)  \right)+ \mathcal{O}\left( 1/S_{j}^2\right) \label{MyRSAndMuGAsymptoticEquation} \end{equation}
where $\sigma^2_{u} \approx \frac{\alpha_{g}+2}{(\alpha_{g} \mu_{g}+1)(\alpha_{g} (1-\mu_{g})+1) }$ and $\sigma^2_{e}$ behaves as $2/S_{j}$, are logit-scale linear mixed regression model group and noise variances respectively.  With unconstrained priors, leaving $\lvec{R}$ confounding, it is uncertain that $\hat{\mu}_{g}$ would converge.

 Table~\ref{Parametertable} summarizes our parameters and prior choices.  We next describe choices in MCMC algorithm that enables investigation of this model, including these constrained priors.

 \begin{table} \caption{Summary of parameters involved in our beta-regression model.}

 \centering
  \fbox{\begin{tabular}{c|l|p{8.5cm}}
  &Prior & Purpose \\ \hline
 $\mu_{\mbox{\tiny{All}}}$ & $\mbox{Beta}(1,1)$ & Overall mean expression of all crosses\\ 
 $\alpha_{\mbox{\tiny{All}}}$ & $\mbox{Gamma}(.1,.1)$ & Beta dispersion of $\vecmu$ about $\mu_{\mbox{\tiny{All}}}$ \\
 $\vecmu$ & $\mbox{Beta} | \mu_{\mbox{\tiny{All}}}, \alpha_{\mbox{\tiny{All}}}$ & Mean parameters $\mu_{g}$ for each cross $g \in \{1, \ldots, G\}$\\
 $\vecalpha$ & $\mbox{Gamma}(1,1)$ & Dispersion parameters $\alpha_{g}$ for each cross $g$\\ \hline
 $\veca$ & $\sum_{k} a_{k}=0$ & In ``\weightbiasedcoin'' model (later Section \ref{Section3:AlleleEffectModeling}) for $\vecmu$, additive effects \\
 $\vecm$ & $\sum_{k} m_{k} = 0$ & In \weightbiasedcoin model, \parentoforigin effects \\ \hline
 $\vecS$ & $\mbox{Gamma}(\xi_{S},\chi_{S})$ & For \tissuegenes $j \in \{1,2,\ldots J\}$, models the \textbf{precision} of that tissue   \\
 $\xi_{S}, \chi_{S}$ & $\mbox{Gamma}(.1,.1)$ & Group dispersion hyper-parameters for the $\vecS$ parameters\\
 $\vecR$ & $\sum_{j} \mbox{logit} R_{j} = 0$ & Models the \textbf{bias} of \tissuegene $j$ toward maternal or paternal expression relative to a true individual-level cell count $P_{i}$ \\
 $\matEta$ & $\sum_{g} \etagj = 0$ & Measures differential in \textbf{precision} about $\vecS$ where certain crosses $g$ are less
  accurate for tissue $j$  \\
  $\tau^2_{\eta}$ & $\mbox{Inv-Chi-Square}(1)$ & Extra dispersion parameter for $\matEta$ in addition to constraint.\\
  $u_{R}$ & $\mbox{Gamma}(1,1)$ & Additional dispersion parameter for $R_{j}$ to have marginal $R_{j}^{u_{R}-1}(1-R_{j})^{u_R-1}$ proportional marginal prior density
  \\ \hline
  
  $\vecP$ & $\mbox{Beta}|\vecmu,\vecalpha$ & For each individual $i \in \{1,\ldots n\}$, models mean maternal 
	  cell active proportion \\ \hline
  $\boldsymbol{Y}^{obs}$ & & Observed portion of $n \times J$ data matrix $\boldsymbol{Y}^{\mbox{\tiny{all}}}$ of proportions of maternal gene expression \\
  $\boldsymbol{Y}^{\mbox{\tiny{miss}}}$ &$\mbox{Beta}|\vecP, \vecS, \vecR, \matEta$ & Missing or unobservable data in $\boldsymbol{Y}^{\mbox{\tiny{all}}}$ that is imputed
  \end{tabular}}  
	  \label{Parametertable}
  \end{table}  \subsection{Parameter Estimation through Slice Sampling}\label{SliceSamplingSection}
 If $\boldsymbol{\Theta}$ represents the unknown model parameters, 
 the product of our model-based likelihood, $P(\matY| \vecP, \boldsymbol{\Theta}) \times P(\vecP| \boldsymbol{\Theta})$
   and the priors, $p(\boldsymbol{\Theta})$, gives an unnormalized function proportional to the desired posterior $\mathcal{P}(\boldsymbol{\Theta}| \matY)$:
\begin{equation} \begin{split} 
   \mathcal{P}( \vecmu, \vecalpha, \mu_{\mbox{\tiny{All}}},  \vecS, & \vecR, \matEta, \xi_{S}, \chi_{S}, u_{R}, \tau^2_{\eta}, \vecP | \matY) 
	    \propto  \mbox{Prior}(\vecmu, \vecalpha,\ldots, 
	   u_{R}, \tau^2_{\eta}) \times \\
	& \prod_{i=1}^{N} \mbox{\small{Be}}\mbox{\small{ta}}(P_{i}; \mu_{g} \alpha_{g}+1, (1-\mu_{g}) \alpha_{g} +1) \times \\
	& \mbox{\hspace{2mm}}
	\prod_{j=1}^{J}\mbox{\small{Beta}}(Y_{ij} ; P_{i} R_{j} S_{j} e^{\etagj} +1, (1-P_{i})(1-R_{j}) S_{j} e^{\etagj}+1)
\mbox{.} \end{split} 
\label{PosteriorFormula} \end{equation}  
 We use the technique of slice-sampling~\citep{Roberts2012,Neal2003} to achieve MCMC random draws from the posterior distribution $\mathcal{P}(\vecmu, \vecalpha, \ldots | \matY)$.  Let group $g \in \{1, \ldots, G\}$, and let $\backslash g$ be $\mathcal{G} - \{g\}$, or the set $\{1,2, \ldots, g-1, g+1, \ldots, G\}$.  In slice-sampling, a new parameter $\alpha_{j}^{(t+1)}$ is drawn given current $\alpha_{g}^{(t)}$ as well as fixed $\vecmu^{(t)}, \vecalpha^{(t)}_{\mbox{\footnotesize{$\backslash g$}}}, \vecP^{(t)}$, by considering the marginal, one dimensional function: $f_{\alpha_{g}}(\alpha_{g} | \vecalpha^{(t)}_{\mbox{\footnotesize{$\backslash g$}}}, \vecmu^{(t)}, \vecP^{(t)})$, which is the marginal unnormalized posterior at a given point $\alpha_{g}^{(t+1)}$ with other parameters fixed. In this case, given $\vecP$ and $\vecmu$, $\vecalpha$ is independent of the $\mu_{\mbox{\tiny{All}}}$,  $\alpha_{\mbox{\tiny{All}}}$, $\vecS$,  $\vecR$, $\matEta$, $\xi_{S}$, $\chi_{S}$, $u_{R}$, $\tau^2_{\eta}$ parameters.  

  To perform slice-sampling, first a uniform value $U_{1}$ is drawn from 
 \begin{equation}U_{1} \sim \mbox{Uniform}\left(0, f_{\alpha_{g}}(\alpha_{g}^{(t)} | \vecalpha^{(t)}_{\mbox{\footnotesize{$\backslash g$}}}, \vecmu^{(t)}, \vecP^{(t)})\right)\mbox{.}\label{Uniform1DrawSliceSampling} \end{equation}  Then the slice sampler seeks left and right from point $\alpha_{g}^{(t)}$ to find the points $\alpha^{-}$ below and $\alpha^{+}$ above s.t. $f_{\alpha_{g}}(\alpha^{-} | \vecalpha^{(t)}_{\mbox{\footnotesize{$\backslash g$}}}, \vecmu^{(t)}, \vecP^{(t)})$ $=$ $f_{\alpha_{g}}(\alpha^{+} | \vecalpha^{(t)}_{\mbox{\footnotesize{$\backslash g$}}}, \vecmu^{(t)}, \vecP^{(t)})$ $=$ $U_{1}$.  To complete the draw, a second uniform is drawn from the interval $[\alpha_{g}^{-}, \alpha_{g}^{+}]$.  This draw serves as the new draw of $\alpha_{g}^{(t+1)}$.  Looping this procedure for all parameters serves as the MCMC scheme.  The algorithmic order can be understood to reflect the order of unobserved parameters in Table~\ref{Parametertable}.
 
 This procedure is well-known and has an advantage over Metropolis-Hastings procedures in not requiring tuning or a proposal density.  Slice-sampling is the backbone to the all-purpose Bayesian integration software ``JAGS''~\citep{Plummer2003, Murphy2007}.  For our purposes, we reimplemented this process explicitly in R SHLIB-compiled C~\citep{RSHLIBRef} for two reasons.  First, we sought to gain efficiency in coding $f(\cdot|\cdot)$ densities by separating out conditionally independent parameters: for instance, conditional on $\vecP$, the posteriors for $\vecS$ and $\vecmu$ are independent.  Efficiencies could be gained by using BLAS~\citep{BLASRef} vector functions: conditional on $\matY$, the posterior for $P_{i}$ is proportional to the vector dot products $\sum_{j} S_{j} e^{\etagij}R_{j} \log_{e}Y_{ij}$ and $\sum_{j} S_{j} e^{\etagij}(1-R_{j}) \log_{e}(1-Y_{ij})$, which are tabulated faster with block memory management.  
 
 But, more crucially, we needed to modify the slice-sampler method to allow for multivariate constrained prior distributions on $\vecR$ and $\matEta$.  Recalling that we require $\sum_{g} \etagj = 0$, we draw new samples for $\etagj^{(t+1)}$ not by slice-sampling along the univariate density $f_{\etagj}(\etagj| \vecR^{(t)}, \vecS^{(t)}, \matEtaNotj^{(t)}, \vecP^{(t)}, \matY)$ but along the full multivariate posterior density as defined along a vector, $\vecEtaj^{(t)}(g,\Delta)$, which is perturbed by a $\frac{\Delta}{G-1}( G e_{g} - \vec{1})$, where $e_{g}$ is the standard $g$th basis vector in $\mathbb{R}^{G}$ and $\vec{1}$ is a G-length vector of 1's.  In other words: 
 \begin{equation} \vecEtaj^{(t)}(g,\Delta) = \left(\eta_{1,j}^{(t)}-\frac{\Delta}{G-1}, \eta_{2,j}^{(t)}-\frac{\Delta}{G-1}, \ldots, \etagj^{(t)} + \Delta, \ldots, \eta_{G,j}^{(t)}- \frac{\Delta}{G-1}\right) \label{RDelta} \end{equation}
 In this case our notation considers perturbation of a column vector $\vecEtaj \in \mathbb{R}^{G}$ which represents all $\eta_{g,j}$ values for a single \tissuegene $j$ fixed and the crosses $g \in \{1,2, \ldots, G\}$.  
 
 To perturb $\vecR$, first define a logit basis reparameterization $L_{j} \equiv \log_{e} \frac{R_{j}}{1-R_{j}}$, and    take a perturbation, $\frac{\Delta}{J-1}(Je_{j} - \lvec{1})$, in the $\lvec{L}$ logit space.
   
 It is important to note a crucial difference in the densities used in slice-sampling of $\matEta$ versus sampling of $\vecR(\lvec{L})$.  One cannot slice-sample over arbitrary reparameterizations without changing measure.  For instance, in a univariate case, if $e^{\omega} = \theta$, then, if $q_{\theta}(\theta)$ was the target density then $q_{\omega}(\omega) = q_{\theta}(e^{\omega}) \times  e^{\omega} $.  It would be inappropriate to find the two $\omega^{+}, \omega^{-}$ such that $f_{\theta}(e^{\omega^{+}}) = f_{\theta}(e^{\omega^{-}}) =\mbox{Unif}(0, f_{\theta}(e^{\omega^{(t-1)}}))$ and then sample $\omega^{(t)} = \mbox{Unif}(\omega^{-}, \omega^{+})$.  One would see a dramatically downward-biased distribution for $\theta(\omega)$.  
For transformations $\theta = \omega-5$ or $\theta = 4 \omega$, slice-sampling from $f_{\theta}(\theta(\omega))$ or $f_{\theta}(\theta)$ produces acceptably distributed samples because the linear transformation does not change the relative measure along $d\theta$.

 In the multivariate setting, consider parameterization $\omega \in \mathbb{R}^{p'}$ that transforms linearly onto the desired parameter space $\theta \in \mathbb{R}^{p}$ with $p' \leq p$,  such as $\theta = \mathbf{A} \omega + \vec{b}$.  Let $\mathbf{A}_{1}$ be the top $p' \times p'$ square matrix of $\mathbf{A}$ and $\mathbf{A}_{2}$ be the remainder $(p-p')\times p'$ rectangular matrix of $\mathbf{A}$.  If matrix $\mathbf{A}_{1}$ is invertible, then $\boldsymbol{\theta}_{1:p'}$ or the first $p'$ coefficients of $\boldsymbol{\theta}$ determine $\boldsymbol{\theta}_{(p'+1):p}$ which are the last $p-p'$ coefficients of $\boldsymbol{\theta}$.  
  
	 If $q_{\boldsymbol{\theta}}(\boldsymbol{\theta})$ were a target density of $f_{\boldsymbol{\theta}}(\boldsymbol{\theta})$ of unknown integration constant, subject to constraints, then $q_{\boldsymbol{\theta}}(\boldsymbol{\theta}) = \frac{f_{\boldsymbol{\theta}}(\boldsymbol{\theta}_{1:p'}, \boldsymbol{\theta}_{(p'+1):p})}{\int f_{\boldsymbol{\theta}}(\boldsymbol{\theta}_{1:p'}', \boldsymbol{\theta}_{(p'+1):p}') d\theta_{1}' \ldots d\theta_{p'}'}$. 
	But also,
	 \begin{equation} q_{\boldsymbol{\omega}}(\boldsymbol{\omega}) = \frac{ f_{\boldsymbol{\theta}}( \mathbf{A}_{1} \boldsymbol{\omega} + \vec{b}_{1:p'}, \mathbf{A}_{2} \boldsymbol{\omega}+ \vec{b}_{(p'+1):p}) |\mathbf{A}_{1}|}{ \int f_{\boldsymbol{\theta}}(\mathbf{A}_{1} \boldsymbol{\omega}'+\vec{b}_{1:p'}, \mathbf{A}_{2} \boldsymbol{\omega}'+ \vec{b}_{(p'+1):p})  |\mathbf{A}_{1} |d\omega_1' \ldots d\omega_p'}. \label{qOmegaTransformedEquation} \end{equation}   In Equation~\ref{qOmegaTransformedEquation}, the Jacobian $|\mathbf{A}_{1}|$ is part of the integration constant.   Slice-sampling will automatically evaluate this constant.  
	
	In the case of $\vecEtaj$, the rotation $\mathbf{A}_{1}$ is the matrix of $-1/(G-1) \vec{1}\vec{1}^{T} + \frac{G}{G-1}\mathbf{I}$ in $\mathbb{R}^{G-1}$.  If $\mathbf{A}_{2} = 1/(G-1) \vec{1}$ and $\vec{b} = \vec{0}$ then the reparameterized space $\vecEtaj' \in \mathbb{R}^{G-1}$ generates $\vecEtaj = \{ \mathbf{A}_{1} \vecEtaj', \mathbf{A}_{2} \vecEtaj'\}$ and we need not change measure.

  In the case of $\vecR$, which includes a nonlinear transformation $\vec{L} \rightarrow \vecR$, the Jacobian matrix is of more consequence;
 \begin{equation} \left| \begin{array}{cccc} \frac{\partial}{\partial L_{1}} \frac{ e^{L_{1}}}{1+e^{L_{1}}} & 
   \frac{\partial}{\partial L_{2}} \frac{ e^{L_{1}}}{1+e^{L_{1}}} & \ldots & \frac{\partial}{\partial L_{J}} \frac{ e^{L_{1}}}{1+e^{L_{1}}} \\
   \frac{\partial}{\partial L_{1}} \frac{ e^{L_{2}}}{1+e^{L_{2}}} & \frac{\partial}{\partial L_{2}} \frac{ e^{L_{2}}}{1+e^{L_{2}}} & \ldots & \ldots \\
     \ldots & \ldots & \ldots & \ldots \\ 
    \frac{\partial}{\partial L_{J}} \frac{ e^{L_{1}}}{1+e^{L_{1}}}
    & \ldots & \ldots  & \frac{\partial}{\partial L_{J}} \frac{ e^{L_{J}}}{1+e^{L_{J}}} \end{array} \right| = \prod_{j=1}^{J} R_{j}(1-R_{j})\mbox{ .} \label{RjPriorNonTrivJacobian} \end{equation}
		The change of variables from $\vecR$ to $\lvec{L}$  measure is
				\begin{equation} f_{\lvec{L}}(\lvec{L}| \matY, \vecS,  \vecP, \matEta, u_{R})=f_{\vecR}(\vecR(\lvec{L}) | \matY,  \vecS, \vecP, \matEta, u_{R}) \prod_{j} R_{j}(L_{j})\left( 1- R_{j}(L_{j}) \right). \label{IdentityOfLandRRatio} \end{equation}
		As with $\vecEtaj$, a linear transformation from $\lvec{L}' \in \mathbb{R}^{J-1}$, such that $\lvec{L}_{1:(J-1)} = \mathbf{A}_{1} \vec{L}'$ and $\lvec{L}_{(p'+1):p} = \mathbf{A}_{2} \lvec{L}'$,  maintains the $\sum_{j} L_{j} = 0$ constraint.

\begin{figure}[htbp]\begin{center}
		\includegraphics[width=1.00\textwidth]{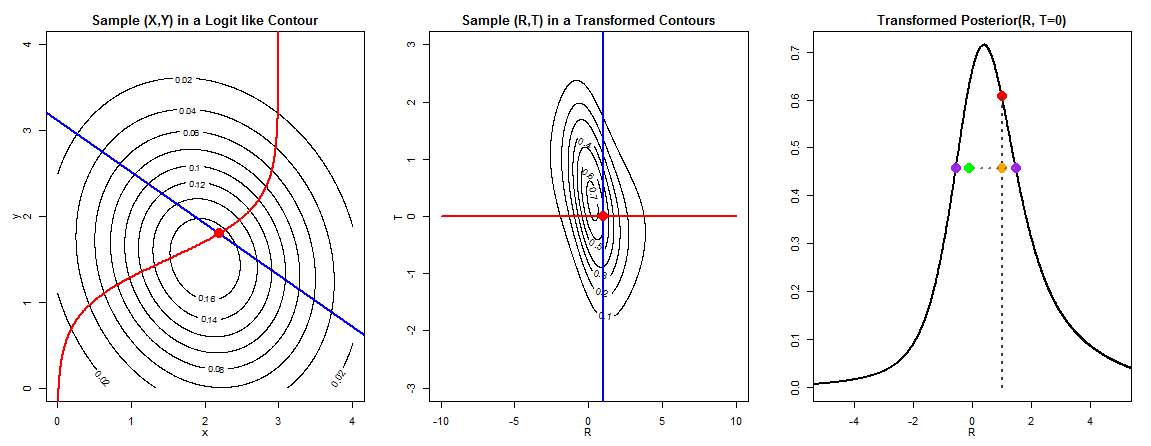} \end{center}
	\caption{A two-dimensional example of the issues faced using a change in variables for slice-sampling.  In the original plot, a Gaussian posterior density in 2 dimensions is considered, but it is desired to sample new $(X,Y)$ along the red curve representing a $(X=\frac{e^{R}}{1+e^{R}}, Y= .3 R+1.5)$ parameterization.  Thus, converting the density of the $(R,T)$ plane involves a Jacobian term, and the observed posterior is warped and non-Gaussian.  Still, slice sampling can be performed for drawing a new $R$ with $T=0$ fixed, where in first a random height of density is chosen underneath the original red dot starting point.  Then a uniform variable is drawn from the band of values for $R$ who have posterior density who have equal or higher density than the selected height.} 
	\label{fig:SkewSamplerExample}
\end{figure}
 
  We note that not all $Y_{ij}$ are observed, sometimes because SNPs do not differentiate between crosses, and in other cases because of  random experimental missingness.  Imputation of the data missing at random is naturally accommodated in a Bayesian setting.  After every iteration, the MCMC sampler draws new samples $Y_{ij}^{\mbox{\tiny{miss}}}$ missing data based upon current estimates of $P_{i}$ and $S_{j}, R_{j}, \etagj$.  Using these new draws of $Y^{\mbox{\tiny{miss}}}$ to make the data complete, samples of $P_{i}, S_{j}, R_{j}, \etagj | \matY^{\mbox{\tiny{complete}}}$ are drawn.  When a gene is entirely excluded  for a certain cross because of identical alleles, we do not impute $Y_{ij}^{\mbox{\tiny{miss}}}$ for that cross, and treat $\etagj$ as essentially non-existent.
 
 Because we vectorize and separate certain operations, computational costs are proportional to $\mathcal{O}(NJ)$ to update all $\vecP, \vecS$ parameters, but $\mathcal{O}(NJ^2)$ to update all $\vecR$ parameters and $\mathcal{O}(JGN)$ to update all $\matEta$ parameters, as well as $\mathcal{O}(G)$ to update $\vecmu, \vecalpha$ parameters.  Thus estimation of $\vecR$ seems to be the most inefficient step in the computation, where most of the computation costs come in calculation of beta values $\frac{ \Gamma(R_{j}P_{i}S_{j} e^{\etagj}+1) \Gamma((1-R_{j})(1-P_{i})S_{j}e^{\etagj}+1)}{\Gamma( (1-R_{j} -P_{j} + 2R_{j}P_{j}) S_{j}e^{\etagj}+1)}$, which do not allow for the efficient separation of parameters.  
 
 Gibbs samples can be used to infer posterior mean, standard-deviation, and quantiles which we use as evidence to simultaneously test biologically relevant hypotheses, such as $\mathcal{H}_{\mu}\mbox{: } \mu_{g} = .5$, $\mathcal{H}_{S}\mbox{: } S_{j_1} = S_{j_2}$, and $\mathcal{H}_{\eta}\mbox{: } \sum_{g}(\etagj-\bar{\eta}_{j})^2 < 10$.  In particular, tests for hypotheses of the form $\mathcal{H}_{\mu_{1}, \mu_{2}}\mbox{: } \mu_{g_{1}} = \mu_{g_{2}}$ are desired, and we hope to use posterior values to both defend and reject hypotheses of this type.

	We can achieve roughly 50K samples in a day, and the Gelman-Rubin~\citep{Gelman1992} convergence measures reported in Section~\ref{Section5:DataAnalysis}  using the complete data are quite low, so we conclude that the chain is optimized sufficiently for the purpose of testing our biological hypotheses.

\subsection{Allele Effect Modeling}\label{Section3:AlleleEffectModeling}
   Previous literature~\citep{Ruvinsky2009, Kambere2009,Wang2010} has suggested  the alleles in \Xce behave in a manner akin to a ``\weightbiasedcoin''.  Consider a cross between mother carrying \Xce allele ``$a$'' and the father with allele ``$b$''.   One could imagine putting a coin together with a mass $M_{a}$ for the heads face, and a mass $M_{b}$ for the tails face.  If a physical model suggested $\mu(a \mbox{ $\times$ } b) = M_{a}/(M_{a} + M_{b})$, we could measure $M_{c}$ in a third allele and predict $\mu(a \mbox{ $\times$ } c) = M_{a}/(M_{a}+M_{c})$.  This model is consistent with a hypothesis that \Xce varies through copy number variation  (CNV).  Longer sequences with more copies attract nucleosomes and transcription factors to attach to the \Xchromosome and deactivate it,  suggesting that longer sequences at \Xce are \textbf{weaker} alleles.

 Now that we have declared multiple inbred strains of mice to carry \Xcea, \Xceb, \Xcec, \ldots alleles, we can implement this model within our framework.    Denote the previous specification, where $\mu_{g}$ are given i.i.d. priors, our ``Independence Assumption'' (IA) model.
After analysis, we can compare IA estimates, where no structure was assumed for $\mu_{g}$, with estimates from this \weightbiasedcoin (WBC) model.   
 
  Consider $d(g)$ be the allele carried by the mother or ``dam'' for cross $g$, and $s(g)$ be the allele carried by the father or ``sire'' inbred-strain for cross $g$.  We model $\mu_{g}$ as a function of additive allele parameters $a_{1}, a_{2}, \ldots, a_{K_a}$ for $K_a$ total alleles and \parentoforigin allele-specific parameters $m_{1}, m_{2}, \ldots, m_{K_m}$ for a total number of \parentoforigin alleles $K_{m}$.  In our case we propose: 
  \begin{equation} \mbox{logit}(\mu_{g}) = (a_{d(g)} + m_{d(g)}) - (a_{s(g)}-m_{s(g)}) \mbox{.} \label{LogitDamCross} \end{equation}
   Having a larger $a_{d(g)}$ dam allele than the $a_{s(g)}$ sire allele pulls $\mu_{g}$ in the direction $\mu_{g} > .5$.  \Parentoforigin effects $m_{d(g)}$ perturb this result based upon whether strains serve as dam or sire.
  
  Again, we impose constraints $\sum_{k=1}^{K_a} a_{k} = 0$ and $\sum_{k=1}^{K_m} m_{k}=0$.  Vector-constrained slice sampling then allows for posterior sampling of these model parameters, along with the previously mentioned $\boldsymbol{\eta}, \vecR$.

 \section{Verifying Method Performance through Simulation}\label{Section4:Simulations}
 Gibbs sampler models are computationally slow, and one can rarely explore all regimes to test a Bayesian estimator.  It would be wrong to draw parameters from prior distributions to simulate datasets; to demonstrate frequency performance of posterior estimates we must use fixed parameter values. 
  We simulate datasets of size $n=200$ individuals, $J=6$ tissue gene measurements, making $G=5$ groups with $40$ samples each.  We fix $\vecmu^{\text{truth}}  = (.25, .45, .5, .65, .75)^{T}$, which is an ordered set of realistic values for the group means,  where adjacent $\mu_{g}$ values are separated by .2, .05, .15, and .1.  We choose \tissuegene precisions $\vecS^{\text{truth}} = (200,200,100,100,50,50)$ and set 
	$\vecR^{\text{truth}}$ such that $\mbox{logit}(\vecR^{\text{truth}}) = (-3,-2,-1,0,1,5)/8$.   Note that the most precise gene-expression values will be downward biased.  We will keep the true secondary precision matrix $\boldsymbol{\eta}^{\text{truth}} = \boldsymbol{0}$, however, the Bayesian estimator will still fit $\boldsymbol{\eta}$ as a free $5 \times 6$ matrix.  We set $\vecalpha^{\text{truth}} = (50,50,50,50,50)^{T}$.  We draw a $P_{i}$ for every individual $i$ based upon group membership $g(i)$, from which $Y_{ij}$ values are sampled, before 66\% of $Y_{ij}$ are removed to represent data missingness (requiring at least one measurement per individual).   

 Our estimation draws 2000 Gibbs samples from the Bayesian Posterior, with fit time approximately 10 minutes, and disk usage approximately 100mb. 
We estimate  posterior mean and credibility intervals for all model parameters.  We ran simulations on the UNC KillDevil Cluster, which allowed roughly 200 cores in parallel,  fitting 1000 replications of all of our simulation experiments (including those described later in this section) in roughly a day.   We compare Bayesian estimates for $\mu_{g}$ to the sample mean of groups $g$:  $\bar{Y}_{g} = \frac{1}{\|Y^{\mbox{\tiny{obs}}}_{ij}\|_{0}}\sum_{\mbox{\tiny{$g(i)=g$}}, \mbox{\tiny{$j$ observed}}} Y_{ij}^{\mbox{\tiny{obs}}}$, where, confidence intervals are based upon the $t$-distribution using standard deviation estimated from the data.  
 
 \begin{table}
\caption{Simulating 1000 datasets from a fixed $\vec{\vecmu}$ vector in model}
\centering
\begin{tabular}{ll|ccccc}
     & $\mbox{True } \mu$ & 0.25 & 0.45 & 0.5 & 0.65 & 0.75\\  \hline 
    Bayes & $\mbox{Mean (Bias)}  \hat{\mu} - \mu$ & -0.003 & -0.001 & -0.001 & 0.001 & 0\\ 
    Method & $\mbox{Mean } \sqrt{(\hat{\mu} - \mu)^2}$ & 0.015 & 0.02 & 0.022 & 0.025 & 0.027\\ 
     & $95$\% $\mbox{HPD Width}$ & 0.056 & 0.063 & 0.063 & 0.061 & 0.056\\ 
     & $95$\%  $\mbox{HPD Coverage}$ & 0.974 & 0.972 & 0.962 & 0.962 & 0.969\\  \hline 
    Sample & $\mbox{Mean (Bias)}  \hat{\mu} - \mu$ & 0.022 & 0.003 & -0.002 & -0.016 & -0.025\\ 
    Mean & $\mbox{Mean } \sqrt{(\hat{\mu} - \mu)^2}$ & 0.026 & 0.015 & 0.016 & 0.022 & 0.028\\ 
     & $95$\% $\mbox{CI Width}$ & 0.054 & 0.061 & 0.06 & 0.056 & 0.051\\ 
     & $95$\%  $\mbox{CI Coverage}$ & 0.619 & 0.957 & 0.933 & 0.786 & 0.487\\ 
  \end{tabular}
  \label{Model1Metthod1}
	\end{table}

    Simulation results in Table~\ref{Model1Metthod1}, show that sample means $\bar{Y}_{g}$ do not estimate the vector $\mu$ well, but that Bayesian credibility intervals have near-95\% coverage of true $\mu_{g}$ values.  For this $n=200$ dataset, with these values of $\vecS$ and $\vec{\vecalpha}$, Bayesian 95\% credibility intervals will be of approximate width 6\%.  This credibility width is sufficient to differentiate almost all of the group $\mu_{g}^{(t)}$.  Only estimates for $\mu_{2} = .45$ and $\mu_{3} = .5$ had overlapping posteriors.  We use a rejection criterion  $2 \times \mbox{min}( \mathcal{P}(\mu_{3}> \mu_{2}, \mu_{2} > \mu_{3}))  < .05$.  That is, we identify two $\mu_{j}$ parameters as different if the two-sided tail posterior probability that their order is reversed is less than $.05$.   For the comparison of $\mu_{2}, \mu_{3}$, this test has a power of 60.4\%.  Comparisons of group means for $\mu_{g} - \mu_{g'} \geq .1$ have near 100\% power.

    Arguably, the sample mean, $\bar{Y}_{g}$, is not meant to estimate $\mu_{g}$, which is a quantity of the  model.  For a fairer comparison,  consider the target of a population gene-expression mean $\bar{Y}^{\mbox{\tiny{pop}}}_{g}$ of group $g$. 
This would be the population average gene expression of the six genes for all possible mice of cross $g$.  If we simulate 100K individuals or more from the model, we can calculate a satisfactory approximation $\bar{Y}^{\mbox{\tiny{pop}}}_{g} = \frac{1}{J \times 100K} \sum_{\mbox{\tiny{100K simulated pups $i'$}}} Y_{i'j}$.  The sample mean $\bar{Y}_{g}$ of 40 pups should be an estimate for $\bar{Y}_{g}^{\mbox{\tiny{pop}}}$.  
In Table~\ref{Model1Table2Means}, we see that, in estimation of observable  $\bar{Y}_{g}^{\mbox{\tiny{pop}}}$ that the Bayesian $95$\% 
credibility intervals tend to be conservative and over-cover $\bar{Y}_{g}^{\mbox{\tiny{pop}}}$.    
The sample mean's nominal $95$\% confidence intervals offer only $90$\% 
coverage for $\mu_{3}, \mu_{4}, \mu_{5}$.  
The bootstrap has wider bootstrap intervals than the Bayesian credibility intervals, but have less coverage.  The Bayesian credibility intervals are .01 narrower, 
but have over-conservative $97$\% 
coverage of $\bar{Y}_{g}^{\mbox{\tiny{pop}}}$.

	\begin{table}	\caption{Simulation performance in estimating population simulated mean.}
	\centering
	\begin{tabular}{ll|ccccc}
    $\mbox{ }$& $\mbox{True } \bar{Y}_{g}^{\mbox{\tiny{pop}}}$ & 0.273 & 0.453 & 0.498 & 0.633 & 0.725\\
     \hline 
    Bayes & $\mbox{Mean }  \hat{\bar{Y}}_{g} - \bar{Y}_{g}^{\mbox{\tiny{pop}}}$ & 0.004 & 0 & 0 & -0.002 & -0.004\\ 
     & $\mbox{Mean } \sqrt{\hat{\bar{Y}}_{g} - \bar{Y}_{g}^{\mbox{\tiny{pop}}})^2}$ & 0.013 & 0.012 & 0.013 & 0.013 & 0.014\\ 
     & $95$\% $\mbox{HPD Width}$ & 0.049 & 0.055 & 0.055 & 0.053 & 0.05\\ 
     & $95$\%  $\mbox{HPD Coverage}$ & 0.974 & 0.972 & 0.957 & 0.962 & 0.976\\
     \hline 
    Sample & $\mbox{Mean }  \hat{\bar{Y}}_{g} - \bar{Y}_{g}^{\mbox{\tiny{pop}}}$ & 0 & 0 & 0 & 0.001 & 0\\  
    Mean & $\mbox{Mean } \sqrt{(\hat{\bar{Y}}_{g} - \bar{Y}_{g}^{\mbox{\tiny{pop}}})^2}$ & 0 & 0 & 0 & 0 & 0\\  
     & $95$\% $\mbox{CI Width}$ & 0.047 & 0.052 & 0.052 & 0.047 & 0.042\\
     & $95$\%  $\mbox{CI Coverage}$ & 0.914 & 0.936 & 0.894 & 0.9 & 0.899\\
     \hline 
    Bootstrap & $\mbox{Mean }  \hat{\bar{Y}}_{g} - \bar{Y}_{g}^{\mbox{\tiny{pop}}}$ & -0.001 & 0 & 0 & 0.001 & 0\\   
     & $\mbox{Mean } \sqrt{(\hat{\bar{Y}}_{g} - \bar{Y}_{g}^{\mbox{\tiny{pop}}})^2}$ & 0.014 & 0.015 & 0.016 & 0.015 & 0.012\\  
     & $95$\% $\mbox{BI Width}$ & 0.054 & 0.061 & 0.06 & 0.056 & 0.051\\  
     & $95$\%  $\mbox{BI Coverage}$ & 0.939 & 0.955 & 0.936 & 0.937 & 0.951
  \end{tabular}
  \label{Model1Table2Means}
  \end{table}

From the Bayesian model, we can also generate an estimate for $\bar{Y}_{g}^{\mbox{\tiny{pop}}}$ by simulating data from the posterior predictive distribution for future $Y_{i'j}$.  For each step $(t)$ in the Gibbs Sampler, we can simulate 100K mice from the model using the current state of  $(\vecmu^{(t)}, \vecalpha^{(t)})$ and calculate expected gene expressions using  $(\vecS^{(t)}, \vecR^{(t)}, \matEta^{(t)} )$, taking the average $\bar{Y}_{g}^{\mbox{\tiny{sim $(t)$}}}$.  These samples, $\bar{Y}_{g}^{\mbox{\tiny{sim $(t)$}}}$, treated as draws from the posterior distribution for $\bar{Y}^{\mbox{\tiny{pop}}}_{g}$.

We compare estimates for $\bar{Y}^{\mbox{\tiny{pop}}}$ using the Bayesian posterior predictive distribution, the sample mean, as well as the bootstrap mean including 95\% bootstrap intervals, in Table~\ref{Model1Table2Means}.  

 Table~\ref{TableAllParamModel1Table} presents average performance of the Bayesian model in estimating the fixed $\vecmu^{\text{truth}}, \vecalpha^{\text{truth}}, \vecS^{\text{truth}}, \vecS^{\text{truth}}, \matEta^{\text{truth}}$ as well as fitting randomly simulated draws for $\vecP$.  Parameter $\vecalpha$ tends to be under-covered, as it is a hierarchy parameter meant to describe variation in $\vecP$, which are unobserved latent data in the model.  On this simulation, where $\vec{\mu}$ was fixed, the hyper parameters hyper parameters such as $\tau^2_{\eta}, \alpha_{\mbox{\tiny{all}}}$ are undefined.
 
 \begin{table}
\caption{Bayesian estimator performance measuring other model parameters}
 \centering
\begin{tabular}{l|cccccc} 
   & $ \vec{\alpha} $ & $ \vec{\mu} $ & $ \mathbf{P} $ & $ \vec{S} $ & $ \vec{R} $ & $ \boldsymbol{\eta} $  \\
   \hline $\mbox{$\sqrt{ }$Average Squared Bias }  \hat{\boldsymbol{\theta}}$  &  $ 6.994 $ & $ 0.001 $ & $ 0.002 $ & $ 31.091 $ & $ 0.001 $ & $ 0.026 $\\
     $\mbox{$\sqrt{ }$MSE for }  \hat{\boldsymbol{\theta}}$  &  $ 13.383 $ & $ 0.022 $ & $ 0.046 $ & $ 52.129 $ & $ 0.019 $ & $ 0.591 $\\
     $\mbox{Mean HPD Width }$  &  $ 53.007 $ & $ 0.06 $ & $ 0.157 $ & $ 189.364 $ & $ 0.039 $ & $ 2.482 $\\
     $\mbox{Mean HPD Coverage}$  &  $ 0.882 $ & $ 0.968 $ & $ 0.931 $ & $ 0.969 $ & $ 0.927 $ & $ 0.964 $\\
     \end{tabular}
          \label{TableAllParamModel1Table}
 \end{table}
  \subsection{Simulation from an Alternate Model}
  Arguably, the beta-distribution model may not reflect the true mechanism generating the data.  We simulate data an alternate linear mixed-effects model on the logit scale:
  \begin{equation} \mbox{logit} (Y_{ij}) = \mu_{g(i)} + \delta_{i} + \mbox{Bias}_{j} + N(0, \sigma^2_{j}) \mbox{.} \label{AlternateSimulationModel}
  \end{equation}
 In the model~\ref{AlternateSimulationModel}, observed $Y_{ij}$ are Gaussian on the logit scale, with a parameter $\mu_{g}$ to represent cross means, $\delta_{i}$ is the $i^{\mbox{\tiny{th}}}$ individual's average deviance from the population mean, and each \tissuegene $j$ supplies a different sized bias and measurement noise.  We use $\mbox{Bias}_{j} = \log_{e} \frac{R_{j}^{\text{truth}}}{1-R_{j}^{\text{truth}}}$ using values of $R_{j}^{\text{truth}}$ from the simulation above, and use $\sigma^2_{j} = 1/S_{j}^{\text{truth}}$.  The $\delta_{i}$ are simulated from a $N(0, 1/\alpha_{g}^{\text{truth}}=1/50)$ distribution.

We fit our original Bayesian model to data generated from the alternate model, and then simulate data from the posterior predictive distribution to give the Bayesian model's estimate for $\bar{Y}^{\mbox{\tiny{pop}}}_{g}$.  In Table~\ref{Model2Table2}, we compare, as in Table~\ref{Model1Table2Means}, performance of the three estimators in this case  where the model assumption for the Bayesian method is incorrect.   The Bayesian method has the shortest intervals and yet is conservative, while the sample mean and bootstrap confidence intervals can under-cover.
  
  \begin{table}
	 \caption{Simulation estimating $\hat{\bar{Y}}_{g}^{\mbox{\tiny{pop}}}$ from  alternate linear mixed-effects}	\centering
\begin{tabular}{ll|ccccc}
     & $\mbox{True } \hat{\bar{Y}}_{g}^{\mbox{\tiny{pop}}}$ & 0.256 & 0.451 & 0.499 & 0.645 & 0.744\\ \hline 
    Bayes & $\mbox{Mean }  \hat{\bar{Y}}_{g} - \bar{Y}_{g}^{\mbox{\tiny{pop}}}$ & 0.001 & 0 & 0 & -0.001 & -0.001\\
     & $\mbox{Mean } \sqrt{(\hat{\bar{Y}}_{g} - \bar{Y}_{g}^{\mbox{\tiny{pop}}})^2}$ & 0.005 & 0.006 & 0.006 & 0.006 & 0.005\\
     & $95$\% $\mbox{HPD Width}$ & 0.025 & 0.031 & 0.031 & 0.029 & 0.025\\
     & $95$\%  $\mbox{HPD Coverage}$ & 0.99 & 0.985 & 0.982 & 0.986 & 0.986\\ \hline 
    Sample & $\mbox{Mean }  \hat{\bar{Y}}_{g} - \bar{Y}_{g}^{\mbox{\tiny{pop}}}$ & 0 & 0 & -0.001 & 0 & 0\\
    Mean & $\mbox{Mean } \sqrt{(\hat{\bar{Y}}_{g} - \bar{Y}_{g}^{\mbox{\tiny{pop}}})^2}$ & 0.008 & 0.009 & 0.01 & 0.008 & 0.007\\
     & $95$\% $\mbox{CI Width}$ & 0.031 & 0.038 & 0.038 & 0.034 & 0.028\\
     & $95$\%  $\mbox{CI Coverage}$ & 0.947 & 0.956 & 0.946 & 0.951 & 0.948\\ \hline 
    Bootstrap & $\mbox{Mean }  \hat{\bar{Y}}_{g} - \bar{Y}_{g}^{\mbox{\tiny{pop}}}$ & 0 & 0 & 0 & 0 & 0\\
     & $\mbox{Mean } \sqrt{(\hat{\bar{Y}}_{g} - \bar{Y}_{g}^{\mbox{\tiny{pop}}})^2}$ & 0.009 & 0.011 & 0.011 & 0.009 & 0.008\\
     & $95$\% $\mbox{BI Width}$ & 0.033 & 0.041 & 0.042 & 0.037 & 0.03\\
     & $95$\%  $\mbox{BI Coverage}$ & 0.918 & 0.935 & 0.939 & 0.945 & 0.945\\
  \end{tabular}\label{Model2Table2}
  \end{table}
   \subsection{``Weight-Biased Coin'' Model}
   We simulate data from the ``weight-biased coin'' model (WBC) to show our performance in measuring $\veca$.  We fix 6 alleles with additive strengths $\veca^{\text{truth}} = \{-5,-1,0,1,2,3\}^{T}/8$, but there will be no \parentoforigin effect, $\vecm^{\text{truth}} = \{0,0,0,0,0,0\}^{T}$.   We produce $n=320$ individuals per simulation, which will be in groups of size $40$ from a random selection of 8 crosses from the $6 \times 6-6=30$ possible crosses.  We simulate a study design that guarantees each allele is sampled at least once, but do not guarantee that an allele is featured both as a dam and a sire.  In the real experiment, allele membership was not known or anticipated until the crosses were performed.  We use simulation to judge whether a random selection of crosses of this type can properly differentiate $\alpha_{k}, \alpha_{k'}$.
   We use the same $\vecR^{\text{truth}}, \vecS^{\text{truth}}, \matEta^{\text{truth}}, \vecalpha^{\text{truth}}$ from before and verify that confidence intervals cover with desired accuracy.

	\begin{table}
	
	\caption{Performance of $\hat{a}_{k}$ estimates in Bayesian WBC model} 
	\centering
\begin{tabular}{ll|cccccc}
     & $\mbox{True $a_{k}$}$ & -0.625 & -0.125 & 0 & 0.125 & 0.25 & 0.375\\
     \hline 
    Bayes & $\mbox{Mean }  \hat{a}_{k} - a_{k}$ & -0.003 & 0.006 & 0 & -0.001 & 0 & -0.043\\
    
     & $\mbox{Mean } \sqrt{(\hat{a}_{k} - a_{k})^2}$ & 0.108 & 0.077 & 0.075 & 0.077 & 0.08 & 0.09\\
    
     & $95$\% $\mbox{HPD Width}$ & 0.624 & 0.608 & 0.611 & 0.609 & 0.602 & 0.616\\
    
     & $95$\%  $\mbox{HPD Coverage}$ & 0.998 & 0.998 & 0.999 & 1 & 0.999 & 0.997\\
    
  \end{tabular} 
  \label{BadModel3Table}
   \end{table}
  
   Table~\ref{BadModel3Table} shows that, while coverage is a proper 95\%, the confidence intervals are  quite wide, approximately $.6$ or $\pm .3$ for every allele effect.  This is disheartening; adjacent allele effects are rarely distinguished in the posterior.   HPDs for $\veca$ are affected by uncertainty for $m_{k}$ reciprocal effects, which is large when a strain has not served at least once as both dam and sire.  For this reason, we retest in Table~\ref{EffectsTable4} in the same simulation framework, but where $\vecm$ is set to zero.   This produces narrower HPDs by 2/3 allowing adjacent $a_{j}, a_{j'}$  to be distinguished.  
	
	 \Parentoforigin effects do, however, appear to be present in the real data, indicating nonzero $\vecm$.   We conclude that a study  would need more than 8 crosses to study these alleles and demonstrate in Table~\ref{Goodmodel5} that 16 of 30 crosses is sufficient.   In the real dataset we collected data from 12 out of 20 allele combinations  We used our simulations to encourage experimenters to perform additional reciprocal crosses. 
  
     \begin{table}
  \caption{Repeat of Table~\ref{BadModel3Table} where $\vecm = 0$ is assumed}
		\centering
\begin{tabular}{ll|cccccc}
     & $\mbox{True $a_{k}$}$ & -0.625 & -0.125 & 0 & 0.125 & 0.25 & 0.375\\
     \hline 
    Bayes & $\mbox{Mean }  \hat{a}_{k} - a_{k}$ & 0 & 0 & 0 & 0 & 0 & 0\\
    
     & $\mbox{Mean } \sqrt{(\hat{a}_{k} - a_{k})^2}$ & 0.044 & 0.042 & 0.043 & 0.041 & 0.042 & 0.04\\
    
     & $95$\% $\mbox{HPD Width}$ & 0.176 & 0.172 & 0.171 & 0.17 & 0.173 & 0.172\\
    
     & $95$\%  $\mbox{HPD Coverage}$ & 0.965 & 0.964 & 0.956 & 0.961 & 0.96 & 0.964\\
  \end{tabular}
 \label{EffectsTable4}
 \end{table}

\begin{table}
\caption{Simulation where $\vecm$ is estimated featuring 16 randomly sampled crosses}
  \centering
  \begin{tabular}{ll|cccccc}
     & $\mbox{True $a_{k}$}$ & -0.625 & -0.125 & 0 & 0.125 & 0.25 & 0.375\\
     \hline 
    Bayes & $\mbox{Mean }  \hat{a}_{k} - a_{k}$ & -0.002 & 0.001 & 0.002 & -0.001 & 0 & 0.001\\
    
     & $\mbox{Mean } \sqrt{(\hat{a}_{k} - a_{k})^2}$ & 0.029 & 0.029 & 0.029 & 0.028 & 0.028 & 0.028\\
    
     & $95$\% $\mbox{HPD Width}$ & 0.124 & 0.119 & 0.119 & 0.118 & 0.119 & 0.121\\
    
     & $95$\%  $\mbox{HPD Coverage}$ & 0.967 & 0.965 & 0.954 & 0.959 & 0.971 & 0.968\\
    
  \end{tabular}

  \label{Goodmodel5}
  \end{table}
   
 \section{Data Analysis}\label{Section5:DataAnalysis}
    	The~\cite{JohnCalaway2013} dataset comprises 660 mice, with 24 \tissuegene combinations, for brain, liver, and kidney.  Missingness is such that for 15840 possible measurements, only 2393 could be observed; thus 84.9\% of potential $\matY$ is unobserved.  
In addition, 34  mice, with 10 brain-derived \RNAseq gene measurements are provided from a set of \WSB, \PWK, 
\CAST wild-derived \FOne crosses. 
These crosses were measured on a different set of 10 genes and no pyrosequencing measurements were taken.  Using both our IA and  WBC  models for $\vecmu$, we implement Gibbs samplers using the observed $\matY$  to draw from the posterior of $\vecP$, $\vecmu$, and $\vecS$.  In Figure~\ref{fig:ShrinkPlot}, we demonstrate how the posterior mean estimates for $P_{i}$ compare to observed $Y_{ij}$ values, using a shrinkage diagram~\citep{Effron1977}.  Since estimates of $P_{i}$ are influenced by $\mu_{g}$ values,  fitted values, $\hat{P}_{i}$,  are different from the unweighted arithmetic mean, $\sum_{j} Y_{ij}/J_{i}$, of the observed gene expressions.  Gene measurements with larger $S_{j}$ values carry higher weight in the estimate of $P_{i}$. 

 For three MCMC chains of length 7500 computed in parallel over the course of a few hours, with i.i.d. $\mbox{Beta}(5,5)$-distributed starting values, the maximum 95\% quantile Gelman-Rubin convergence diagnostic (GRD) is 1.11 for the 32 $\vec{\mu}$ parameters with a multivariate potential scale reduction factor (PSRF) of 1.08 and a mean 5-lag autocorrelation .267~\citep{Gelman1992, Brooks1998}.  In contrast, the maximum 95\% quantile GRD is 1.06 for the 34 $\vecS$ parameters with $\mbox{Gamma}(10)/10$ distributed starting values, and a multivariate PSRF of 1.04, but the mean 5-lag autocorrelation is a much slower .784.  We conclude that mixing of the chains is sufficient for analysis of this dataset.
 
\begin{figure}[htbp]
	\centering
	\includegraphics[width=1.0\textwidth]{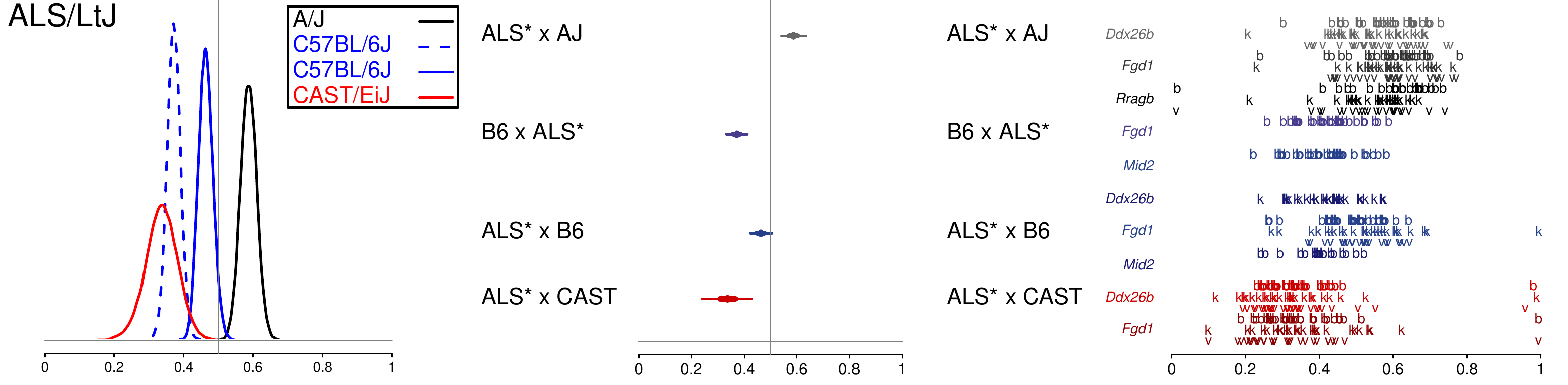}
	\caption{Left: for four crosses using strain \ALS, we plot posterior density for $\mu_{g}$ in terms of active 
	  \ALS.   Blue represents crosses with known \Xceb carriers (dotted line for the reciprocal cross where \BSix served as maternal strain), black for \Xcea carriers, and red for \Xcec.   Center: we plot posterior 95\% credibility intervals for $\mu_{g}$ of these crosses.   The posterior suggests it is highly unlikely that $\mu_{g}$ for the \ALS $\times$ \CAST cross is greater than .5, with mean .34.  The posterior for \ALS$\times$\AJ cross is concentrated above $.5$, with mean .59.  The \ALS$\times$\BSix cross carries  posterior weight over $\mu_{g} = .5$.  The reciprocal \BSix$\times$\ALS cross does not, suggesting significant \parentoforigin bias.  
				On the right are observed gene expressions $Y_{ij}$ used to estimate $\mu_{g}$.  Points are labeled ``b'', ``k'', ``v'' for brain, kidney and liver respectively.}
	\label{fig:Fig2Model9SpecimenALS}
\end{figure}

 The most important scientific objectives are accomplished by estimation of $\vecmu_{g}$.   Most strains were anticipated to carry \Xceb, and so we used three pieces of evidence to converge this suspicion.  We established these candidates carried alleles stronger than \Xcea by showing that $\mu_{g} > .5$ when these candidates served as dam when crossed with known \Xcea carriers, we established these candidates carried alleles weaker than \Xcec by showing that $\mu_{g} < .5$ when they served as dam when crossed with \Xcec carriers, and finally we tried to establish that $\mu_{g}$ was close to $.5$ when these candidates were crossed with known \Xceb carriers.   For crosses these candidates served as sire, the ordering of $\mu_{g}$ is reversed.  We rejected undetermined candidates as allele members when crosses with \preexisting allele holders demonstrated $\mu_{g} \neq .5$.  We used a criterion of $ 2\times \mbox{min}( \mathcal{P}(\mu_{g} > .5 | \mathbf{Y}), \mathcal{P}(\mu_{g} < .5| \mathbf{Y}))$, that is, twice the smallest tail posterior probability on either side of $.5$.   An example of posterior estimation for $\mu_{g}$ in crosses with \ALS is given in Figure~\ref{fig:Fig2Model9SpecimenALS}, and a summary table of crosses of interest is in Table~\ref{Table:NewTableForMeansOfAnalysis}.
These give estimates for $\mu_{g}$ and confidence measures, for crosses of the unknown strains with known ``$\mbox{\emph{Xce}}^{a}$'', 
		``$\mbox{\emph{Xce}}^{b}$'', ``$\mbox{\emph{Xce}}^{c}$'' carriers.   When multiple crosses between two alleles have been conducted, we report $p_{g} = 2 \mbox{min}(P(\mu_{g} > .5),P(\mu_{g} < .5))$ for the lowest posterior value among those crosses.
		
		When 
    multiple crosses, including reciprocals, against an allele carrier were performed,
    we present the estimates with the smallest $p_{g}$.  A ``*'' suggests that we have made
    a scientific decision to associate this strain with this allele.  
    **: \ALS  was tentatively called a ``$\mbox{\emph{Xce}}^{b}$'' since the \ALS$\times$\BSix cross
    had $\mu_{g} = 0.46$ $(0.43, 0.51)$ with $p_{g} = 0.092$ for $30$ animals, despite the reciprocal cross showing
    such a strong  parent-of-origin effect (also for $30$ animals).

     Operating on this principle we confirmed  that  \LEWES, \PERA, \TIRANO, \SJL, \WLA, \WSB, and \ZALENDE carry \Xceb.  Tentatively, we believe \ALS to be a \Xceb carrier, since it is not of type \Xcea or \Xcec.  While \BSix$\times$\ALS cross showed bias from $.5$, the \ALS$\times$\BSix seemed much closer to a cross between identical carriers. 
		\PWK appears to have an allele with strength near that of the \Xceb  allele, although previous literature hypothesized that it contains a \Xcec  as does \CAST.  	 
	 \begin{table} 
  \caption{
   Estimate $\hat{\mu}_{g}$,
    tail probability ($p_{g}$) against the hypothesis $\mathcal{H}_{0}: \mu_{g} = .5$.}  
		
	\centering
     
    \begin{tabular}{c|cc|cc|cc} & 
      \multicolumn{2}{c}{$\mbox{\emph{Xce}}^{a}$} &  \multicolumn{2}{c}{$\mbox{\emph{Xce}}^{b}$}  &
      \multicolumn{2}{c}{$\mbox{\emph{Xce}}^{c}$}  \\
     & $\mu_{g}$ & $p_{g}$  &  $\mu_{g}$ & $p_{g}$ & $\mu_{g}$ & $p_{g}$   \\ \hline
   \ALS 
     & $\begin{array}{c} 0.59 \\ (0.54, 0.63) \end{array}$ & $0.003$ 
     & $\begin{array}{c} 0.37 \\ (0.33, 0.41) \end{array}$ & $\mbox{**}0.003$ 
     & $\begin{array}{c} 0.34 \\ (0.25, 0.43) \end{array}$ & $0.002$ \\ \hline
   \SJL 
     & $\begin{array}{c} 0.57 \\ (0.54, 0.61) \end{array}$ & $0.003$ 
     & $\begin{array}{c} 0.48 \\ (0.45, 0.51) \end{array}$ & $\mbox{*}0.142$ 
     & $\begin{array}{c} 0.31 \\ (0.27, 0.35) \end{array}$ & $<.001$ \\ \hline
   \LEWES 
     & $\begin{array}{c} 0.6 \\ (0.56, 0.64) \end{array}$ & $0.003$ 
     & $\begin{array}{c} 0.49 \\ (0.44, 0.53) \end{array}$ & $\mbox{*}0.683$
     &  &  \\ \hline
   \WLA 
     & $\begin{array}{c} 0.55 \\ (0.51, 0.59) \end{array}$ & $0.009$ 
     & $\begin{array}{c} 0.48 \\ (0.44, 0.53) \end{array}$ & $\mbox{*}0.45$
     &  &  \\ \hline
   \WSB 
     & $\begin{array}{c} 0.6 \\ (0.52, 0.67) \end{array}$ & $0.022$ 
     & $\begin{array}{c} 0.47 \\ (0.37, 0.57) \end{array}$ & $\mbox{*}0.545$ 
     & $\begin{array}{c} 0.25 \\ (0.17, 0.33) \end{array}$ & $0.006$ \\ \hline
   \PWK 
     & $\begin{array}{c} 0.67 \\ (0.57, 0.76) \end{array}$ & $0.008$ 
     & $\begin{array}{c} 0.61 \\ (0.54, 0.67) \end{array}$ & $0.007$ 
     & $\begin{array}{c} 0.21 \\ (0.15, 0.28) \end{array}$ & $0.005$ \\ \hline
   \TIRANO 
     & $\begin{array}{c} 0.61 \\ (0.56, 0.66) \end{array}$ & $<.001$ 
     & $\begin{array}{c} 0.51 \\ (0.47, 0.55) \end{array}$ & $\mbox{*}0.511$
     &  &  \\ \hline
   \ZALENDE 
     & $\begin{array}{c} 0.6 \\ (0.56, 0.63) \end{array}$ & $<.001$ 
     & $\begin{array}{c} 0.56 \\ (0.5, 0.61) \end{array}$ & $\mbox{*}0.076$
     &  &  \\ \hline
   \PERA
     &  &  
     & $\begin{array}{c} 0.54 \\ (0.49, 0.59) \end{array}$ & $\mbox{*}0.1$
     &  &  \\ \hline
   \PANCEVO 
     & $\begin{array}{c} 0.46 \\ (0.41, 0.5) \end{array}$ & $0.082$ 
     & $\begin{array}{c} 0.41 \\ (0.37, 0.44) \end{array}$ & $0.003$
     &  &  
   \\
   \hline
   \end{tabular} 
    
   \label{Table:NewTableForMeansOfAnalysis}
   \end{table} 	
	
     \subsection{Analysis of Weight-biased Coin Model}
   Table~\ref{Table:AdditiveMaternalAlleleEffectsFittedAugust2013FFix} displays logit-scale estimated  allele effects based upon our decisions for carriers of 
	\Xcea, \Xceb, \Xcec, as well as yet-unidentified placeholders 
	for \PWK ($\mbox{\emph{Xce}}^{p}$), and \PANCEVO ($\mbox{\emph{Xce}}^{v}$). 
	We observed that \parentoforigin effect is not necessarily attached to \emph{Xce} allele, so we separated strains into potential  \parentoforigin carriers based upon phylogenetic relationships between strains.  We model nonzero $m_{k}$ \parentoforigin only for candidates that serve both as dam and sire, else we fix $m_{k}=0$.  
	.
	   
 \begin{table}
  \caption{Logit-scale estimated allele effects}  
    \begin{tabular}{c|c}  $\mbox{Allele}$ &
    $\begin{array}{c} \mbox{Additive} \\ \veca  \end{array}$
    \\  \hline
       $\mbox{\emph{Xce}}^{a}$ & -0.36(-0.52, -0.18) \\  
       $\mbox{\emph{Xce}}^{b}$ & -0.01(-0.19, 0.19) \\  
       $\mbox{\emph{Xce}}^{c}$ & 1.08(0.47, 1.67) \\  
       $\mbox{\emph{Xce}}^{p}$ & -0.23(-0.42, -0.03) \\  
       $\mbox{\emph{Xce}}^{v}$ & -0.67(-0.88, -0.45) \\
  \end{tabular}  $\mbox{, }$
    \begin{tabular}{c|c}  $\mbox{Maternal}$ & $\vecm$ 
    \\  \hline
      $\mbox{\AJ/\OneTwoNine}$ & 0.09 (-0.05, 0.24) \\  
      $\mbox{\ALS}$ & -0.07 (-0.5, 0.32) \\  
      $\mbox{\BSix}$ & 0.46 (0.25, 0.67) \\  
      $\mbox{\CAST}$ & 0.19 (-0.43, 0.73) \\  
      $\mbox{\LEWES}$ & 0.09 (-0.08, 0.27) \\  
      $\mbox{\PWK}$ & 0.05 (-0.24, 0.36) \\  
      $\mbox{\WSB}$ & -0.81 (-1.54, -0.08) \\
  \end{tabular}
  
   \label{Table:AdditiveMaternalAlleleEffectsFittedAugust2013FFix}
   \end{table} 	
	 The WBC model measures \BSix to have a large .46 \parentoforigin effect $m_{\mbox{\tiny{\BSix}}}$, and $m_{\mbox{\tiny{\WSB}}}$ for \WSB is large and negative.  Wild-derived \emph{mus musculus} representative \PWK appears to carry an additive allele with effect strength between \Xcea and \Xceb.  Results for the \weightbiasedcoin (WBC) model were not considered essential to the \cite{JohnCalaway2013} paper, 	but this model can be incorporated with genetic sequence experiments to search for \parentoforigin loci.

\subsection{Comparison of \WeightBiasedCoin versus Independence Assumption}
Comparing posterior means $\hat{\mu}_{g}$ from the WBC and the original IA model for $\mu_{g}$, the mean $\| \hat{\mu}_{g}^{\mbox{\tiny{WBC}}} - \hat{\mu}_{g}^{\mbox{\tiny{IA}}}\|_{2}$ is .12 and the correlation $\mbox{Corr}(\hat{\mu}_{g}^{\mbox{\tiny{WBC}}}, \hat{\mu}_{g}^{\mbox{\tiny{IA}}})$ is  $.63$.  The cross ``HF'' which is a \WSB$\times$\CAST cross measured using \RNAseq shows the maximum difference with $\hat{\mu}_{g}^{IA} = 0.42\mbox{ }(.297,\mbox{ }.55)$ and 
$\hat{\mu}_{g}^{WBC} = 0.16225\mbox{ }(.053,\mbox{ }0.289)$.

\begin{figure}[htbp]
	\centering
		\includegraphics[width=0.80\textwidth]{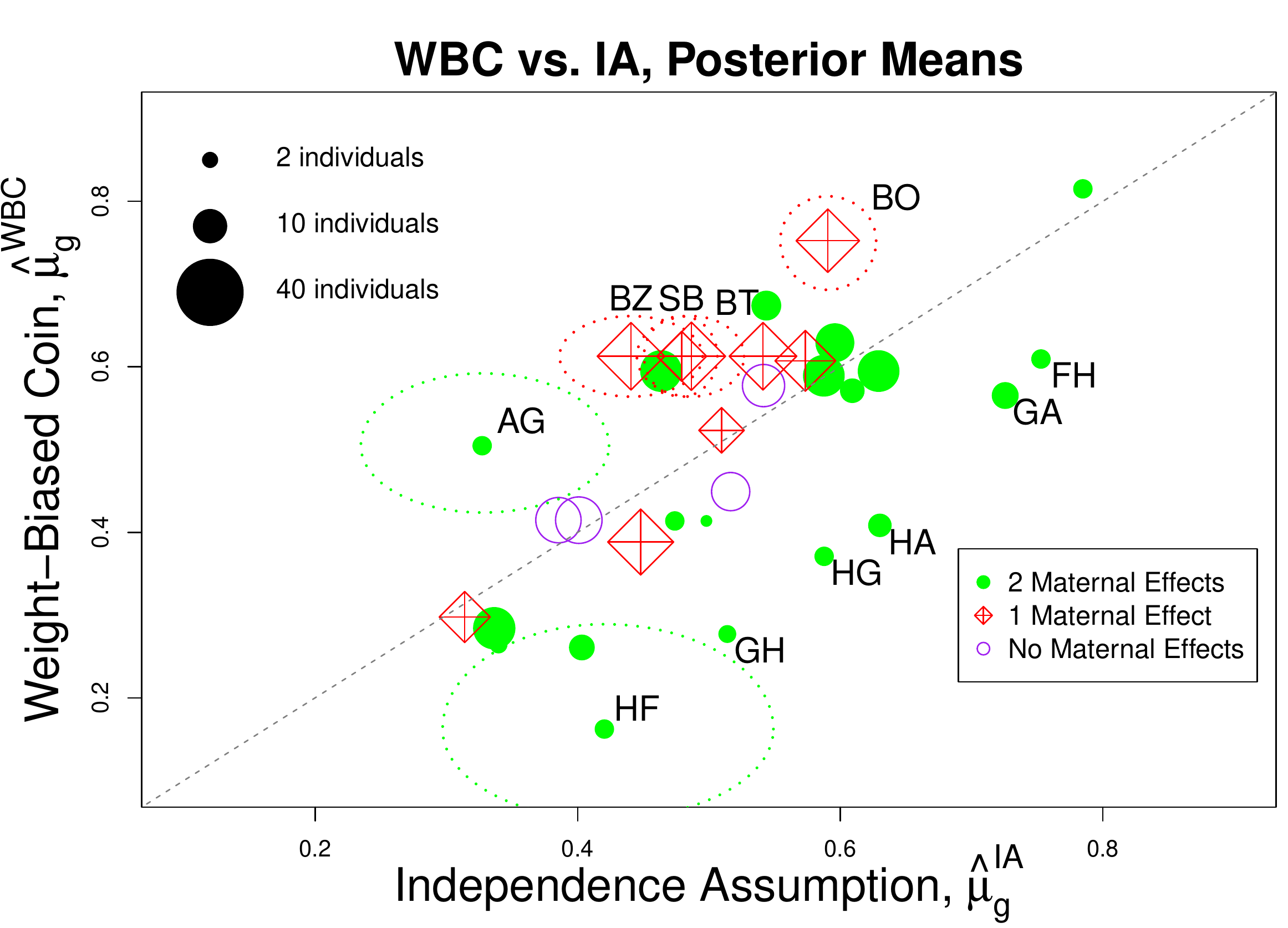}
	\caption{Comparison of posterior means $\hat{\mu}^{\mbox{\tiny{IA}}}$ on x-axis and $\hat{\mu}^{\mbox{\tiny{WBC}}}$ on the y-axis.  Size of point represents the number of individuals used in a cross.  Strains not used as both dam and sire could not be measured for a maternal effect $\vec{m}$.  Green points correspond to crosses between two strains that each contribute a $m_{j}$ effect.  Red points included one strain which had fixed $m_{j}=0$ because it is unidentifiable.  Purple points have both strains fixed $m_{j}=0$.  We plot confidence ellipses for the crosses that do not agree between WBC and IA.  There are 6 crosses out of 34 where $95\mbox{\%}$ HPD intervals between the two models do not overlap.}
	\label{fig:TextAsP19s}
\end{figure}

Figure~\ref{fig:TextAsP19s} demonstrates the differences between WBC and IA models.  Points sizes indicate the number of individuals of each cross,  varying in size from 2 to 43.  Larger crosses contribute more to the likelihood, and the WBC model will sacrifice fit for the smaller crosses.  For the WBC model, strains used only as dam or sire  have fixed $m_{k} = 0$.  The green points in Figure~\ref{fig:TextAsP19s} are between strains which both have non-zero $\hat{m}_{k}$, whereas the red points include one $\hat{m}_{k}$ contribution, and the purple points have none.

HPD intervals in the WBC and IA models do not overlap for only 6 crosses: HF (\WSB$\times$\CAST), AG (\AJ$\times$ \PWK), BZ (\BSix$\times$ \ZALENDE), 
  SB (\SJL$\times$ \BSix), BT (\BSix$\times$\TIRANO) and BO (\BSix$\times$\PANCEVO).  Additionally, in crosses GH (\PWK$\times$\WSB), HG (\WSB$\times$\PWK), HA (\WSB$\times$\AJ), GA (\PWK$\times$\AJ), FH (\CAST$\times$\WSB)  the WBC model seems to underfit the observed result, though these crosses are smaller and HPD intervals overlap.

		Wide disparities between $\hat{\mu}^{\mbox{\tiny{WBC}}}$ and $\hat{\mu}^{\mbox{\tiny{IA}}}$ may suggest crosses have been insufficiently sampled, or gene measurements might be biased, \Xce alleles are mislabeled, or that the logit-linear model is insufficient.  But in general, the WBC model successfully predicts $\mu_{g}$.

\subsection{Precision parameters}

Tables~\ref{Table:NewTableForSValuesPyro} and~\ref{Table:NewTableForSValuesRNA} presents estimated $\vecS$ precision parameters for the IA model.  Some \RNAseq gene measurements (taken only in the brain), have similar $S_{j}$ levels as pyrosequencing.  Ideally, we would have liked to measure both pyrosequencing and \RNAseq on the same tissue, same mouse, same cross same gene.  But we were only able to procure \RNAseq from a set of \PWK-\WSB-\CAST crosses sequenced to serve multiple experiments in addition to our own.  In the pyrosequencing measurements, a gene seems to be more precise in the brain than in the kidney or liver.  Estimates for  $\hat{S}_{j}$ range from 300 (suggesting a s.d. of about $\pm .03$ near $P_{i}=.5$) to some \RNAseq measurements that only display $\hat{S}_{j} = 8$ (an s.d. of about $\pm .17$) accuracy.

 Based upon the P\`{o}lya Urn abstraction described earlier, we can interpret these values suggesting multiple hundreds of cells in the brain at the point of \Xinactivation, relative to 50 cells in the kidney or liver.  Measurements of \emph{Xist}, \emph{Hprt1}, \emph{Acs14} measurements appeared in few crosses, and credibility for $S_{j}$ of these genes range from $10^{-13}$ to $300$, meaning that these genes were not measured in enough crosses provide reliable contribution.

 \begin{table}
	\caption{$\vecS$ precisions from pyrosequencing (brain/kidney/liver) measurements}
	\centering
   \begin{tabular}{c|cccc}
   $\mbox{ }$ &  \emph{ Fgd1 } & \emph{ Ddx26b } & \emph{Rragb1} & \emph{Mid2}   \\
    \hline
    brain & $440.6$ & $103.4$ & $123$ & $385.2$ \\
    & ($214.3$, $692.2$) & ($77.1$, $130.6$) & ($70.2$, $182.8$) & ($138.5$, $683.7$)  \\
    \hline
    kidney & $251.8$ & $49.4$ & $120.7$ & $99.3$ \\
    & ($169.6$, $349.3$) & ($32.9$, $66.3$) & ($74.2$, $169.2$) & ($0.04$, $296.2$)  \\
    \hline
    liver & $81.2$ & $31.4$ & $30.8$ & $105.9$ \\
    & ($61.6$, $101$) & ($23.1$, $39.9$) & ($20$, $42.4$) & ($30.2$, $192.4$)   \\ \hline \hline
   &  \emph{ Zfp182 } & \emph{ Xist } & \emph{ Hprt1 } & \emph{ Acs14 }   
    \\
    \hline
    brain & $289.2$ & $25.5$ & $29.5$ & $26.7$ \\
    & ($62.2$, $581.5$) & ($1.6\times10^{-14}$, $135$) & ($1.2\times10^{-15}$, $151.2$) & ($1.3\times10^{-26}$, $139.9$)  \\
    \hline
    kidney & $109.6$ & $30.4$ & $25.5$ & $26.5$ \\
    & ($0.1$, $330$) & ($1.2\times10^{-11}$, $152.2$) & ($4.3\times10^{-18}$, $131.6$) & ($1.4\times10^{-15}$, $137.2$)  \\
    \hline
    liver & $349$ & $28.5$ & $27.3$ & $26.8$ \\
    & ($84.9$, $650.4$) & ($4.3\times10^{-13}$, $143.3$) & ($1.2\times10^{-18}$, $142.6$) & ($4.2\times10^{-16}$, $138.9$)   \\
   \hline  \end{tabular}
\label{Table:NewTableForSValuesPyro}
\end{table}

\begin{table}
\caption{Precision in genes from \RNAseq Measurements (brain only)}
\centering
   \begin{tabular}{ccccc} 
 \emph{Gripap1} & \emph{Zdhhc9} & \emph{Ids} & \emph{Pls3} & \emph{Arhgef9} \\ 
 $232.4$ & $114.7$ & $534.2$ & $175.4$ & $42.1$ \\
 ($102.5$, $376.2$) & ($45.3$, $191.9$) & ($236.1$, $873.8$) & ($73.1$, $303.4$) & ($11.9$, $76$) \\
 \hline
 \emph{Zmym3} & \emph{Sh3bgr1} & \emph{Gprasp1} & \emph{Gnl3l} & \emph{Tspyl5} \\ 
 $256.7$ & $120.1$ & $8.7$ & $126.9$ & $220.1$ \\
 ($108.2$, $434.4$) & ($46.9$, $206.3$) & ($1.1$, $17$) & ($49.1$, $213.1$) & ($85.8$, $389$) \\
 \end{tabular}
 
\label{Table:NewTableForSValuesRNA}
\end{table}

 \section{Discussion and Conclusions}\label{Section6:Discussion}
  We have developed a Bayesian hierarchical  model, robust and powerful enough to confirm our hypotheses, which uses allele-specific gene-expression to estimate 
	whole-body \Xinactivation.  
In simulations, we showed that Bayesian 95\% HPD credibility intervals reliably 
  cover the truth in simulated datasets similar in size and structure to our experimental dataset, 
	and that these intervals improve coverage relative to asymptotic-$t$ and bootstrap-based estimators, 
	even when data is generated from an alternate model.  
Our ``independence-assumption'' (IA) model, where $\mu_{g}$'s are given i.i.d. priors,  
  demonstrated and confirmed allele membership of 10 previously untested strains, 
	which was the critical analysis of the \cite{JohnCalaway2013} experiment.  
Our \weightbiasedcoin (WBC) model, where $\mu_{g}$ is a function of parental strains, 
  including additive allele effects $\veca$ and reciprocal effects $\vecm$, produced 
	similar results to the IA prior, suggesting this scientific model 
	might be used to predict $\mu_{g}$ in future crosses.  
  
  Modeling bias and precision of gene-specific measurements using  free ($\vecS$) and 
   constrained $(\vecR, \matEta)$ parameters, we developed techniques in constrained slice-sampling that required both linear and non-linear transformation of the parameter set.  
	Our model could accommodate allele-specific expression measured both by pyrosequencing and \RNAseq.
	   
Here we must acknowledge limitations in experimental design that led to Bayesian recourse.   
Quantitative trait loci (QTL) are regions within the genome whose variation is highly correlated, 
  and conjectured to be causal, with a phenotype ---  in our case \Xinactivation skewing.  
Typical experiments to locate QTL, such as Advanced Intercrosses~\citep{Darvasi1995}, 
   in which individuals of known genotype are mated,
	 rely on random recombination of the genome.
But precise measurement of \emph{Xce} bias requires replicates in the form of cloned individuals.
\cite{JohnCalaway2013}'s methodology resulted from fortuitous selection of 
  already-sequenced inbred strains that were found to exhibit sequences of 
	similarity and difference within the~\cite{Chadwick2006} interval.  
Establishing that two strains have heavy difference in a sub-region, 
  yet their offspring exhibit no \Xinactivation skewing, gives credence to a belief that 
	the sub-region of difference is not a candidate for \Xce.  
Establishing that two strains, identical at a sub-region, 
  have offspring that exhibit \Xinactivation skewing,  rejects a sub-region 
	in favor of other regions where these two strains do differentiate genetically.
   
 Prior to experiment we did not know all of which crosses to perform, 
  did not know which sub-region we wished to justify, and did not know 
	which strains would differentiate.  
As early strains in the experiment were established as 
  carriers for known alleles ``$a$'', ``$b$'', ``$c$'', we were further 
	motivated to test other strains that continued to break down the candidate region.  
Such experimental choices of new test strains were done both in the 
  light and in the dark of statistical analysis and estimates; as such,
	our stopping rule cannot be claimed to be a sample-independent choice.  
The ``null'' sampling space in this case is ill-posed: we did not randomly sample from a population, 
  but instead chose sub-populations to explore based upon research intuition.  
Sometimes samples were obtained because they were readily available or
  cheaper to obtain. 
But, at other times, crosses were ordered because they were hoped to 
	lead to a new allele. 
Sometimes additional samples were obtained to improve
	statistical confidence on a cross, adding to samples already collected. 
From a frequentist perspective, it is right to be skeptical of analysis 
  on such a dataset.
Since we  cannot map out the decision metric that led to 
  collection of samples, we cannot postulate the alternative sample space for most hypothesis tests,
	and cannot report a $p$-value.
   
Bayesian exchangeability ~\citep{DeFinneti1937, Bernardo1996, Lindley1976, 
  Hewitt1955, Jackman2009, Diaconis1980} often justifies posterior analysis of sequential designs,
	and here we show the sampling mechanism is an ignorable design~\citep{ADAPTIVESample1996}.
Consider indicators $I_{i,1}, I_{i,2}, \ldots, I_{i,G}$ which represent
  whether individual $i$ is in group $g\in \{1,2,\ldots, G\}$, such that $\sum_{g} I_{i,g} = 1$. 
Let $\lvec{I}_{i}$ vector be assigned 
  with a transition probability 
	$f^{i}_{\text{trans}}(\lvec{I}_{i} | Y_{\mbox{\tiny{$< i$}}}, \mathbf{I}_{\mbox{\tiny{$<i$}}})$, 
	which is a multinomial dependent only upon previous observations and sampling choices.
Conditional on $\lvec{I}_{i}, \vecmu, \vecalpha$, 
  the $P_{i}$ are independent from other individuals.
The full posterior, including arbitrary sequential design, is
   \begin{equation}
   \begin{split} \mathcal{P}( \boldsymbol{\Theta} | \mathbf{Y} ) \propto & \prod_{i=1}^{n} f_{\text{trans}}^{i}(\lvec{I}_{i} | Y_{\mbox{\tiny{$< i$}}}, \mathbf{I}_{\mbox{\tiny{$<i$}}}) \times \\ & 
   \mbox{ } \mbox{ } \mbox{ } \mbox{ }  \mbox{ }   \mbox{\small{Beta}}( P_{i}; \sum_{g} I_{i,g}\mu_{g}\alpha_{g}+1, \sum_{g}I_{i,g}(1-\mu_{g})\alpha_{g}+1) \times \\ & 
   \mbox{ } \mbox{ } \mbox{ } \mbox{ }  \mbox{ } \prod_{j} \mbox{\small{Beta}}(Y_{ij}; P_{i} R_{j}S_{j} e^{\etagj} +1, (1-P_{i})(1-R_{j}) S_{j} e^{\etagj} + 1) \\
  &  \times \mbox{Priors}( \mathbf{S}, \mathbf{R}, \boldsymbol{\mu}, \boldsymbol{\alpha}, \boldsymbol{\eta}) \mbox{ . }\end{split}
   \label{YExchangeableSequence}
   \end{equation}

We see from expanded equation ~\ref{YExchangeableSequence} above that because of 
  conditional independence given $P_{i}$, and because all vectors $\lvec{I}_{i}$ 
	are observed and known, then the transition density,  
	$f_{\text{trans}}^{i}(\lvec{I}_{i} | Y_{\mbox{\tiny{$< i$}}}, \mathbf{I}_{\mbox{\tiny{$<i$}}})$, 
	which is only based upon observables, cannot influence the posterior.
Because experimenters had no preference before observations of allele identities 
  for the individuals chosen, we can accept that the stopping rule is proper 
	(though certainly not optimal) in the definition of a proper stopping rule given 
	in ~\cite{Parmigiani2009}.  We would have eventually stopped collecting samples,
	no matter the true value of $\mu_{g}$.
	
We note, per a criticism in~\cite{Lindley1972}, that our $\vecmu$ parameters
  are not actually exchangeable based upon science.
Even if we do not know the value of $\mu_{g}$, we do know that if
  $\mu_{g}, \mu_{g'}, \mu_{g''}$ were a sequence of crosses of a 
	mother strain with unknown allele ``UNK'' to fathers of allele carriers ``a'', ``b'', ``c'', 
	then  there should be a decreasing sequence: 
	$\mu_{g}$ $\geq$ $\mu_{g'}$$\geq$ $\mu_{g''}$. 
In the ``IA'' model, we did not model allele-order hypothesis with our priors and relied 
  upon the data to reveal this sequence.  	The WBC model did enforce this relation, and reached similar 
  estimates of $\hat{\mu}_{g}$.  
     
 It is difficult to statistically argue for the hypothesis $\mu_{g'} = \mu_{g}$, 
   in contrast to proving $\mu_{g'} \neq \mu_{g}$.
   We have relied upon a criterion
	of overlapping posterior.
It is possible that the difference between two cross means, $\mu_{g} - \mu_{g'} = 
  \Delta$, might be small, non-zero, but imperceptible, such as $\Delta \approx .0000001$.  
In which case, we would inevitably decide incorrectly to treat 
  $\mu_{g}, \mu_{g'}$ as fundamentally equivalent crosses between same alleles.
We  assume that alleles 
  differ in effect enough to be statistically observable.

As high-throughput sequencing becomes cheaper, Bayesian algorithms of our design must be 
  rejected in favor of more efficient methods. 	
The Bayesian algorithm has complexity $\mathcal{O}(NJ^2)$, due to Gibbs sampling of
individual $S_{j}$ and $R_{j}$, making 
  it unsuitable for whole-transcriptome datasets, where a complete genome 
	has $J \approx 8000$.
At the time of research, \RNAseq of a single brain-tissue sample might cost \$1,000 and 
  require multiple weeks of setup, sequencing, and bioinformatic analysis,  
  but pyrosequencing a few genes, to greater precision, of the same sample costs $\approx$\$20.  
A fully-analyzed \RNAseq sample might measure an individual's  $P_{i}$ average brain
  expression to a $\pm .1\%$ region of 95\% credibility, but pyrosequencing of three genes would
	leave a $\pm 2\%$ region.
		But to measure cross mean $\mu_{g}$, it is better to have $P_{i}$ measured 
  imprecisely in many individuals in multiple tissues rather than have high precision on 
	any single-tissue $P_{i}$.
  	
We note that point-estimate performances of the sample mean estimator and bootstrap  are 
  nearly identical to the posterior mean.  Were it not for under-coverage in their confidence intervals, and a conceptual disconnect 
  between these procedures and our parametric model, these methods might have served as suitable
	replacements to the Bayesian Gibbs sampling.

To the extent we have introduced techniques more generally applicable 
  to statistics, our constrained 
  priors show promise in other naturally unidentifiable problems with small order $J$.  
We are fortunate to be in a data-setting where a Bayesian estimator is statistically 
  robust and viable, and can use posterior sampling to test a parametric 
	model for its comparable nuances.   

\section{Acknowledgements}

This project was supported by National Institutes of Health (NIH) grants R01GM104125 (ABL, WV), R35GM127000 (WV), and P50MH090338 and P50HG006582 (FPMdV, JDC). We also thank UNC Information Technology Services for computational support. The funders had no role in study design, data collection and analysis, decision to publish, or preparation of this manuscript

  \bibliography{xinactivation-methods_references}
	\end{document}